\documentclass[12pt,preprint]{aastex}

\slugcomment{Accepted to ApJ}

\shorttitle{Molecular outflow survey in the OMC-2/3 region}
\shortauthors{Takahashi et al.}

\begin{document}

\title{Millimeter- and Submillimeter-Wave Observations\\
of the OMC-2/3 Region. III. An Extensive Survey for Molecular Outflows}

\author{SATOKO TAKAHASHI}
\affil{Academia Sinica Institute of Astronomy and Astrophysics, P.O. Box 23-141, Taipei 106, Taiwan; $satoko{\_}t@asiaa.sinica.edu.tw$ \\
Department of Astronomical Science, The Graduate University for Advanced Studies, 
National Astronomical Observatory of Japan, Osawa 2-21-1, Mitaka, Tokyo 181-8588, Japan}

\author{MASAO SAITO}
\affil{ALMA Project Office, National Astronomical Observatory of Japan, Osawa 2-21-1, Mitaka, Tokyo 181-8588, Japan}

\author{NAGAYOSHI OHASHI}
\affil{Academia Sinica Institute of Astronomy and Astrophysics, P.O. Box 23-141, Taipei 106, Taiwan}

\author{NOBUHIKO KUSAKABE}
\affil{Department of Astronomical Science, The Graduate University for Advanced Studies, 
National Astronomical Observatory of Japan, Osawa 2-21-1, Mitaka, Tokyo 181-8588, Japan}

\author{SHIGEHISA TAKAKUWA}
\affil{Academia Sinica Institute of Astronomy and Astrophysics, P.O. Box 23-141, Taipei 106, Taiwan}

\author{YOSHITO SHIMAJIRI}
\affil{Department of Astronomy, School of Science, University of Tokyo, Bunkyo, Tokyo 113-0033, Japan}

\author{MOTOHIDE TAMURA}
\affil{National Astronomical Observatory of Japan, Osawa 2-21-1, Mitaka, 
Tokyo 181-8588, Japan}

\author{and \\
RYOHEI KAWABE}
\affil{Nobeyama Radio Observatory, Nobeyama, Minamimaki,Minamisaku, Nagano,
384-1305, Japan}

\newpage

\begin{abstract}
Using the ASTE 10 m submillimeter telescope and the 1.4 m Infrared Survey Facility (IRSF), 
we performed an extensive outflow survey in the Orion Molecular Cloud -2 and -3 region. 
Our survey, which includes 41 potential star-forming sites, 
has been newly compiled using multi-wavelength data based on millimeter- and submillimeter-continuum observations 
as well as radio continuum observations. 
From the CO (3--2) observations performed with the ASTE 10 m telescope, 
we detected 14 CO molecular outflows, seven of which were newly identified. 
This higher detection rate, as compared to previous CO (1--0) results in the same region, 
suggests that CO (3--2) may be a better outflow tracer.  
Physical properties of these outflows and their possible driving sources were derived. 
Derived parameters were compared with those of CO outflows in low- and high-mass star-forming regions. 
We show that the CO outflow momentum correlates with the bolometric luminosity of the driving source 
and with the envelope mass, regardless of the mass of the driving sources. 
In addition to these CO outflows, seven sources having NIR features suggestive of outflows were also identified.  
\end{abstract}


\section{INTRODUCTION}
	
	CO molecular outflows are ubiquitous phenomena in low- to high-mass star-forming regions 
	(e.g. Bachiller \& Tafalla 1999 for reviews, Zhang et al. 2001, Arce et al.  2007 for reviews). 
	This phenomenon is considered to be directly associated with the main accretion phase in protostars, 
	and is frequently used to identify protostars in dense cores. 
	Hence the CO outflows provide the primary information concerning the mass accretion 
	and ejection processes during the protostellar evolution. 
	Previous single-dish surveys revealed the physical nature of CO outflows toward hundreds of low- to high-mass young stellar objects  
	(e.g., Cabrit \& Bertout 1992, Shepherd \& Churchwell 1996, Bontemps et al. 1996, Beuther et al. 2002, Zhang et al. 2001, 2005). 
	An extensive outflow survey of low-mass objects by Bontemps et al. (1996) found that the momentum flux of the CO 
	outflows is positively correlated with the bolometric luminosity and also positively correlated with the mass of the associated parent cores. 
	Recent high-mass studies also show the above correlations continuing to the high-mass 
	region over several orders of magnitude in mass and bolometric luminosity (e.g., Curchwell 1997, Beuther et al. 2002, Zhang et al. 2005). 
	On the other hand, the systematic studies focused on outflows associated with intermediate-mass 
	protostars ($2~{\leq}~M_{\ast}~{\leq}~8~M_{\odot}$) are still limited, 
	although several candidates have been investigated in detail to date 
	(e.g., Fuente et al. 2005, Takahashi et al. 2006, Beltran et al. 2006, 2007). 
	Studies of the intermediate-mass outflows are important for revealing the evolution of intermediate-mass stars.
	But more importantly, such studies address the question of wether the same basic accretion and outflow processes operate continuously 
	from low- to high- mass stars.
	
	In this paper, we present the results of our extensive outflow survey toward one of the nearest 
	intermediate-mass star-forming regions: The Orion Molecular Cloud -2/3 (OMC-2/3; $d$=450 pc; Genzel \& Stutzki 1989). 
	The goals of this project are to (i)search for outflows in OMC-2/3 with high sensitivity and high spatial resolution, 
	(ii) combine the detected outflow results and the centimeter- to mid-infrared continuum information to identify a driving source of the outflows, 
	and (iii) reveal the structures and the physical properties of each outflow.
	In the OMC-2/3 region, 38 millimeter- and submillimeter dust continuum sources were detected 
	(Chini et al. 1997, Lis et al. 1998, Nielbock et al. 2003). 
	Free-free jets were also detected at 3.6 cm toward 11 of the embedded sources, and eight of 11 sources 
	are associated with the dust continuum sources (Reipurth et al. 1999). 
	Shock-excited H$_2$ knots, which probably trace shock fronts within each flow, 
	are associated with these embedded sources (Yu et al. 1997, 2000, and Stanke et al. 2002). 
	Aso et al. (2000) and Williams et al. (2003) made mapping observations in the CO (1--0) emission and identified nine CO outflows. 
	They found that the outflows in OMC-2/3 are mostly spatially compact 
	(i.e., possess short dynamical time scales of a few $\times~10^4$ yr), and energetic 
	(${\sim}$ several ${\times}~10^{-4}$ M$_{\odot} $km s$^{-1}$). 
	However, previous outflow identifications have ambiguities, 
	because the CO (1--0) emission is often significantly affected by the contamination from the ambient molecular cloud. 
	This made it difficult to identify small and faint outflows and differentiate them from the ambient molecular gas. 
	Furthermore, pre- and proto-stellar candidates in OMC-2/3 are lying along the filament within a few $\times$ 0.1 pc. 
	The crowded core distributions also made it difficult to identify which is the driving source of each outflow. 	
	
	We report results of our high-sensitivity molecular outflow survey in the submillimeter CO (3--2) emission 
	with the ASTE telescope. The ASTE (Atacama Submillimeter Telescope Experiment) telescope equipped 
	with a 345 GHz receiver is one of the most powerful facilities with which to seek molecular outflows over a wide field of views. 
	The ASTE telescope in the ``On-The-Fly'' mapping mode enabled us to observe the entire region of OMC-2/3 with 
	a high sensitivity limit (i.e., corresponding mass of the sensitivity limit is $\sim$10$^{-4}~M_{\odot}$).
	The CO (3--2) emission exclusively traces higher temperature gas ($\geq$ 33 K). 
	Our CO (3--2) results are compared with previous CO (1--0) results.   
	We also show deep $JHK_s$ images taken by the Simultaneous Three-Color InfrarRed Imager for Unbiased Survey camera 
	(SIRIUS) on Infrared Survey Facility (IRSF), which provide us with the distributions of faint reflection nebulae, chains of knots, and jet-like features. 
	Furthermore, we also used the archived 24 $\mu$m data obtained by the infrared camera MIPS (Multiband Imaging Photometer) 
	on the infrared space telescope Spitzer, which are sensitive to hot dusty components with a temperature of 
	${\sim}$ 150 K, and probably trace 100 AU scale innermost envelopes/circumstellar 
	disks associated with the central heating sources.
	
	The details of the observations of the submillimeter CO (3--2) emission 
	and the near infrared $JHKs$-bands are presented in Section 2. 
	In Section 3, we show the results of the outflow survey. 
	Data analysis of the CO (3--2) outflows is shown in Section 4 and properties of the outflow driving sources are described 
	in Section 5. In Section 6, we discuss the physical properties of the CO outflows in OMC-2/3 and compare them with 
	outflows associated with low- to high-mass YSOs in previous studies. 
	Finally, Section 7 summarizes the paper.

\section{OBSERVATIONS AND DATA REDUCTION}
	\subsection{CO (3--2) observations}
	The CO ($J$ = 3--2; 345.795990 GHz) data have been taken with the ASTE 10 m telescope 
	(Ezawa et al. 2004; Kohno et al. 2004) located
	at the Pampa la Bola (altitude = 4800 m), Chile. 
	Observations were remotely made from the ASTE operation room at San Pedoro de Atacama, 
	using the network observation system N-COSMOS3 developed by NAOJ (Kamazaki et al. 2005) 
	during the period of September 2nd to 4th, 2005.
	The half-power beam-width of the ASTE telescope is 22$''$. 
	We used the 345 GHz SIS heterodyne receiver, which had the typical system noise temperature of 175 - 350 K 
	in DSB mode at the observed elevation. 
	The typical atmospheric opacity at 220 GHz was 0.05 toward the zenith during our observations. 
	The temperature scale was determined by the chopper-wheel method, 
	which provides us with the antenna temperature corrected for the atmospheric attenuation. 
	As a backend, we used four sets of 1024 channel auto-correlators, providing us with 
	a frequency resolution of 31.12 KHz that corresponds to 0.27 km s$^{-1}$ at the CO (3--2) frequency.
	The On-The-Fly (OTF) mapping technique (Sawada et al. 2008) was employed
	to cover the whole OMC-2/3 region, $23'.0{\times}13'.3$ (corresponding to 3.1 $\times$ 1.8 pc at 450 pc).  
	The pointing of the telescope was checked every two hours by 5-points scans of the point-like 
	CO (3--2) emission from O Cet. [R.A. (J2000) = 02$^h$ 19$^m$ 20.8$^s$, decl. (J2000) = -02$^{\circ}$ 58$'$ 40.7$''$].
	The pointing errors measured during the observations range from 1$''$.5 to 4$''$.0.
	To convert the antenna temperature to brightness temperature, 
	we adopted the main beam efficiency of 0.55 which was measured by Sawada et al. (2008) 
	during the same observational semester.
	After subtracting linear baselines, the data were convolved 
	by a Gaussian-tapered Bessel function (Magnum, Emerson, \& Greisen 2007) 
	whose FWHM was 14$''$ and resampled onto a 7$''$ grid.
	Since the telescope beam is a Gaussian with a FWHM of 22$''$, 
	the effective FWHM resolution is 26$''$ (corresponding to 0.06 pc). 
	The ``scanning effect'' was minimized by combining scans along R.A. and decl., 
	using the PLAIT algorithm developed by Emerson, D. T. \& Graeve, R. (1988).
	The typical rms noise level was 0.47 K in brightness temperature ($T_b$) with a velocity resolution 
	of 1.08 km s$^{-1}$ after smoothing over four channels. 
	
	\subsection{NIR observations}
	NIR images in the OMC-2/3 region were obtained in November 2004 with the
	near-infrared camera SIRIUS (Simultaneous Three-Color InfrarRed Imager for Unbiased Survey)  
	mounted on the 1.4 m Infrared Survey Facility (IRSF) telescope at the South African Astronomical Observatory, in Sutherland. 
	The camera is equipped with three 1024 $\times$ 1024 pixel HgCdTe (HAWAII) arrays. 
	Two dichroic mirrors enable us to make simultaneous observations at the
	$J~({\lambda}_{c}=1.25~{\mu}\rm{m})$, $H$ (1.63 $\mu$m), and $K_{s}$ (2.14 $\mu$m) bands.
	Details of the camera are provided by Nagashima et al. (1999) and Nagayama et al. (2003).
	The image scale of the array is 0$''$.45 pixel$^{-1}$, providing a field of view of $7'.7{\times}7'.7$ 
	(corresponding to 1.0 $\times$ 1.0 pc).
	Three different field centers, [1]: R.A. = 5$^h$ 35$^m$ 22$^s$, decl. = -5$^{\circ}$ 00$'$ 14.8$''$,
	[2]: R.A. = 5$^h$ 35$^m$ 22$^s$, decl. = -5$^{\circ}$ 6$'$ 44.9$''$, 
	and [3] R.A. = 5$^h$ 35$^m$ 22$^s$, decl = -5$^{\circ}$ 13$'
	$14.8$''$, were set to cover the entire OMC-2/3 region.
	We obtained a total exposure time of 900 seconds in each field. 
	Seeing conditions were $\sim 1''.0~(FWHM)$ in the $J$ band.
	The limiting magnitudes at $J$, $H$ and $K_s$ -band were 19.2, 18.6 and 17.3 mag., respectively. 
	We used the NOAO IRAF software package to reduce the data.
	We applied the standard procedures for near-infrared array image reduction, 
	including dark-current subtraction, sky subtraction, and flat-fielding 
	(see Nakajima et al. 2005 for details of the SIRIUS image processing).
 
\section{RESULTS} 
	\subsection{Potential Star-Forming Regions in OMC-2/3}
	As the first step of the outflow identification described in the next section, 
	we identified individual star-forming sites in the OMC-2/3 region. 
	For this purpose, we compiled 1.3 mm (Chini et al. 1997, Nielbock et al. 2003) 
	and 350 $\mu$m (Lis et al. 1998) dust continuum sources detected 
	in the OMC 2/3 region, finding 38 sites having dust continuum sources 
	detected at either 1.3 mm, 350 $\mu$m, or both. 
	In addition, we also compiled 3.6 cm radio continuum sources observed with the VLA in 
	the OMC 2/3 region (Reipurth et al. 1999), finding 11 radio continuum sources. 
	Note that eight of the 11 radio continuum sources are associated with the dust continuum sources we compiled, 
	In total, 41 sites listed in Table 1 and shown in Figure 1$a$ are identified 
	as potential star-forming sites located in the OMC-2/3 region. 

	Because some of the sites that we identified may be still in the starless or pre-stellar phase, 
	we checked whether 24 $\mu$m sources observed with Spitzer/MIPS are associated 
	with these potential star-forming sites. 
	The mid-infrared 24 $\mu$m emission, presumably tracing hot dust ($T_{\rm{dust}}{\sim}$150 K) around protostars, 
	is a good indicator of ongoing star-formation. Eighteen sites in Table 1 are 
	associated with 24 $\mu$m emission suggestive of ongoing star-forming activities. 
	
	\subsection{Outflow Identification}
	The search for outflows was done based on the CO (3--2) data of the OMC-2/3 region that we obtained. 
	The actual search, however, was limited to the vicinity of the potential star-forming sites listed in Table 1. 
	This is because the purpose of our search is to identify outflows associated with potential star-forming sites 
	and to compare characteristics of outflows in the following sections. 

	In order to identify outflows, we carefully inspected velocity channel maps of the CO (3--2) emission to see 
	whether there is localized blue-shifted or red-shifted emission in the vicinity (within a radius of ${\sim}1'$) of each site in Table 1. 
	We also used our $JHKs$ images to see whether there were any reflection nebulae, 
	jet-like futures, or a chains of knots, which are suggestive of the existence of outflows, 
	in the vicinity of each site in Table 1. 

	With these search procedures, we identified 21 outflow candidates, 
	which were then classified into three categories, ``CLEAR'', ``POSSIBLE'', and ``MARGINAL'' as follows; 

	\begin{description}
	\item[CLEAR:]
	outflows in this category have both localized blue- and red-shifted lobes 
	with clear bipolarity in the CO (3--2) channel maps.

	\item[PROBABLE:]
	outflows in this category have either a localized blue- or red-shifted lobe in the CO (3--2) channel maps. 

	\item[MARGINAL:]
	outflows in this category have neither a localized blue- nor a red-shifted lobes 
	in the CO (3--2) channel maps, but are associated with extended CO (3--2) high-velocity emission 
	and/or NIR features suggestive of an outflow in the $JHKs$ images.
	\end{description}

	With this classification, 10 outflow candidates were classified as CLEAR, 
	four were classified as PROBABLE, and six were classified as MARGINAL. 
	Table 2 shows the summary of the outflow search for each of the potential star-forming sites. 
	Figures 1$b$ and 1$c$ show an overall distribution of the 14 outflows candidates 
	categorized as either CLEAR or PROBABLE. Among these 14 outflow candidates, 
	seven were also previously found in CO (1--0) (Aso et al. 2000, Williams et al. 2003). 
	The other seven outflow candidates were newly identified in our observations. 
	We also detected Spitzer 24 $\mu$m sources toward 12 out of the 14 outflows as the likely outflow-driving candidate.  
	Eight sources having near-infrared features suggestive of outflows were identified.
	In the next section, we describe the results of individual outflow candidates in detail. 

	\subsection{\bf Individual Outflows}
	In this section, we present individual outflows identified in this paper.
	We also present possible outflow candidates which are not associated with any of the 41 potential star-forming sites 
	in Section 3.4. In the outflow identification, we adopt the source name based on the 1.3 mm sources by Chini et al. (1997) 
	and Nielbock et al. (2003). If driving sources have no 1.3 mm sources, we adopt the 350 $\mu$m and/or VLA-3.6 cm 
	source name which were identified by Lis et al. (1998) and Reipurth et al. (1999), respectively. 
	
	\subsubsection{\bf OMC-3 SIMBA Region}
	Five of the 41 potential star-forming sites are located in the northeastern OMC-3 region, 
	called the ``SIMBA'' region (Nielbock et al. 2003). 
	In this region, we newly identified two outflow candidates in CO (3--2); 
	one, categorized as CLEAR, is associated with SIMBA $a$ as shown in Figure 2, and the other, 
	categorized as PROBABLE, is associated with SIMBA $c$ as shown in Figure 3. 
	The later one associated with SIMBA $c$ has only a blue-shifted lobe, 
	in which a faint NIR nebula denoted by [1] in Figure 3$a$ is found. 
	Note that both the blue-shifted lobe and the NIR nebula show elongation in the same direction (northeast--southwest). 
	The elongated CO and NIR structures would support the presence of the outflow associated with SIMBA $c$.
	Although there is red-shifted CO (3--2) emission to the west of SIMBA $c$, 
	which could be a counterpart of the blue-shifted lobe, 
	it is difficult to conclude that this red-shifted emission is a part of the outflow associated with SIMBA $c$,  
	because the orientation of the red-shifted emission (east--west) is very different from that of the blue-shifted lobe, 
	and also because the red-shifted emission is not clearly isolated from the ambient cloud. 
	Both SIMBA $a$ and SIMBA $c$ are associated with Spitzer 24 $\mu$m sources 
	(SP 053530-045848 and 053528-045838, respectively) as shown in Figure 2$b$, 
	and they could be the driving sources of these outflows.

	\subsubsection{\bf OMC-3 Region}
	Figures \ref{f4}, \ref{f5}, and \ref{f6} show results of our outflow survey in the the OMC-3 region 
	(except the SIMBA region). 14 of 41 potential star-forming sites are located in this region, 
	and at least six 1.3 mm sources have a class 0-type SEDs with a small ratio of $L_{bol}/L_{smm}< 200$ (Chini et al. 1997).
	We identified four CO (3--2) outflows categorized as CLEAR in this region. 
	
	$MMS~2; (CLEAR)$-- 	We detected a CO (3--2) outflow along the east-west direction toward MMS 2 as 
	shown in Figure \ref{f4}$b$. The blue- and red-shifted components overlap, 
	suggesting that the axis of this flow is close to the plane of the sky.
	This east-west outflow has also been identified in the CO (1--0) emission (Williams et al. 2003, Aso et al. 2000). 
	However, the previous CO (1--0) studies had difficulty identifying which sites of MMS 1-4 the CO outflow is associated with. 
	Because only MMS 2 is associated with a significant 24 $\mu$m Spitzer source (SP 053518-050045), 
	this CO (3--2) outflow, presumably driven by the 24 $\mu$m Spitzer source, 
	is most probably associated with MMS 2 rather than MMS1, MMS 3 or MMS 4. 
	In the NIR image shown in Figure 4$a$, there is a jet-like feature and a knot to the west of the Spitzer source 
	(denoted by [3] and [4], respectively). 
	These two features were also detected in NIR H$_2$ emission (Yu et al. 1997, Stanke et al. 2002, Tsujimoto et al. 2004). 
	Both of them are elongated in the same direction and in particular the reflection nebula denoted by [3] is spatially coincident with the blue-shifted lobe.
	An extended NIR feature was also detected in the eastern part of the MMS 2 along the outflow axis (denoted by [2]).  
	Our $JHK_s$ image suggests that the NIR feature is a reflection nebula from the isolated star-forming site 
	rather than knots from MMS 2 (Section 3.4), although previous NIR H$_2$ observations suggested that 
	the origin of the NIR feature is the outflow from MMS 2. 

	$MMS~5; (CLEAR)$-- 
	There is a compact east-west outflow with a clear bipolarity centered on MMS 5, as shown in Figure \ref{f4}$b$. 
	This east-west outflow has also been identified in the CO (1--0) emission (Williams et al. 2003, Aso et al. 2000). 
	A 24 $\mu$m source (SP053522-050115), associated with MMS 5, is located between the blue- and red-shifted lobes, 
	suggesting that this 24 $\mu$m source is most probably the driving source of the outflow. 
	The infrared image in Figure \ref{f4}$a$ shows that several knots, denoted by [5], 
	are associated with the blue-shifted lobe. These knots were also detected in infrared H$_2$ emission 
	(Yu et al. 1997 and Stanke et al. 2002).
	
	$MMS~6; (MARGINAL)$-- 
	MMS 6 is the brightest 1.3 mm continuum source in the OMC-2/3 region (Chini et al. 1997). 
	We did not find any clear signature of CO (3--2) outflows (see Figure \ref{f4}$b$), 
	as also indicated by previous CO (1--0) observations (Williams et al. 2003, Matthews et al. 2005), 
	although there is extended high-velocity red-shifted emission to the southeast of MMS 6. 
	Careful inspection of the near-infrared image in Figure \ref{f4}$c$, however, 
	shows that there are two faint $Ks$ emissions north and south of MMS 6. 
	These two northern and southern sources are also detected at 24 $\mu$m 
	(SP 053524-050130 and SP 053524-050140, respectively). 
	More importantly, the near-infrared image shows a faint jet-like feature denoted by [6] 
	to the east of the northern infrared source. 
	In addition, a 3.6 cm free-free emission (i.e., VLA 3) is also associated with the northern infrared source. 
	These results suggest that there may be an east-west outflow associated with the northern infrared source. 
	We note that previous studies (Yu et al. 1997, Stanke et al. 2002) considered 
	the two infrared sources north and south of MMS 6 to be knots due to a jet, 
	although we consider them to be protostars because they are more point-like and are detected even at 24 $\mu$m.
 
	$MMS~7; (CLEAR)$-- Figure \ref{f5} shows a molecular outflow associated with MMS 7, which
	is elongated in the east-west direction. 
	While the outflow shows a clear bipolarity, the red-shifted lobe is much more elongated than the blue-shifted lobe.
	The connected trajectory of the CO red-shifted emission peaks, denoted as a black solid curve in Figure \ref{f5}$b$, shows a
	wiggled structure over one pc to the west of MMS 7, possibly suggesting a precessing motion.
	This east-west outflow has also been identified in the CO (1--0) observations 
	by Aso et al. (2000), Williams et al. (2003), and Takahashi et al. (2006). 
	The projected maximum velocity of this outflow is relatively small, 
	and the NIR reflection nebula (Haro-5a/6a; Haro 1953) associated with the outflow extends to both east and west. 
	These suggest that the outflow is almost on the plane of the sky.
	In addition, there is a one-sided optical jet on the western side (Reipurth, Bally, \& Devine 1997; HH 294) 
	and a compact 3.6 cm free-free jet associated with the central star (Reipurth et al. 1999, 2004), 
	both of which have roughly the same position angle as that of the CO (3--2) flow.
	We also detected faint knots along the western part of the CO outflow as denoted by [7] in Figure \ref{f5}$a$. 
	We note that the CO outflow associated with MMS 7 has a very large collimation factor, 15.4, for blue-shifted component, 
	which is significantly larger than the others, 0.9--4.7
	\footnote{The axis ratio ($R_{\rm{maj}}/R_{\rm{min}}$) was adopted the maximum size of major- and minor- axes measured by eyes.}. 
	We do not understand how to make such a large collimation factor of CO outflows, 
	but this kind of CO outflow might be related to optical jets on a parsec scale (Reipurth et al. 1997).  

	In the northeast of the MMS 7, there is an embedded protostellar candidate, 
	MMS 7-NE, which has been identified by Takahashi et al. (2006), 
	although the source is not included in the 41 potential star-forming sites.  
	A possible compact bipolar outflow associated with MMS 7-NE was detected 
	in the interferometric CO (1--0) observations with a 0.05 pc scale (Takahashi et al. 2006).  
	The CO (3--2) bipolar outflow associated with the MMS 7-NE was 
	not detected probably due to the insufficient spatial resolution and potential contamination 
	of the MMS 7 outflow emission. 
	
	$MMS~9; (CLEAR)$ -- 
	A prominent CO (3--2) outflow elongated from east to west was detected as shown in Figure \ref{f6}$b$. 
	This outflow has also been identified by CO (1--0) observations (Aso et al. 2000, Williams et al. 2003). 
	The outflow shows clear bipolarity centered on MMS 9. The blue- and red-shifted lobes are east and west of MMS 9. 
	An additional red-shifted lobe was also detected overlapping part of the blue-shifted lobe. 
	The maximum size of the outflow was measured at up to one pc in channel maps. 
	A 24 $\mu$m emission (SP 053526-050546) is associated with MMS 9, 
	and is presumably the driving source of the outflow, 
	although there is another 24 $\mu$m source (SP 053526-050558) located at 10$''$ south of MMS 9. 
	In the $JHKs$ image (see Figure \ref{f6}$a$), we detected a near-infrared feature denoted by [9]  
	located to the east of the 24 $\mu$m source (SP 053526-050546). 
	In addition, there appear bow-shock-like futures as denoted by [10] and [11] in Figure \ref{f6}$a$, 
	which seem to be associated with the termination of the blue-shifted (or red-shifted) CO knots 
	presented in Figure \ref{f6}$b$ [i]. 
	It is interesting to note that the velocity width of the blue-shifted emission abruptly increases 
	at the position of the bow-shock-like futures as shown in the position-velocity diagram (see Figure \ref{f9}). 
		
	Figure \ref{f6}$b$ shows that there are two other outflow candidates in this region. 
	One is the blue-shifted lobe elongated to the northeast denoted by [ii]
	from the position of the 24 $\mu$m sources (SP 053526-050546 and SP 053526-050558). 
	At the northeast end of the blue-shifted lobe, there is a faint bow-shock-like feature 
	detected in the $JHKs$ image shown in Figure \ref{f6}$a$ (denoted by [8]). 
	It is not clear whether this blue-shifted lobe is a part of the outflow described above, 
	or another outflow having no clear red-shifted lobe. 
	In the following discussion, we consider that this blue-shifted lobe is a part of the outflow associated with MMS 9.
	The other outflow candidate is the extended reflection nebula, located at $\sim$15$''$ west 
	of MMS 10 (denoted as [12] in Figure \ref{f6}$a$). There is an infrared point source identified as J05353153-0505473 in the the 2MASS catalog 
	to the east of the reflection nebula. 
	Even though the source sits inside the red-shifted outflow from MMS 9, it is probably not related to the high-velocity emission. 
	This is because the source already appears in the $J$-band as a point source, 
	suggesting that it  is more evolved than other potential star-forming sites 
	
	\subsubsection{\bf OMC-2 Region}
	The OMC-2 region contains embedded clusters with approximately 100 stellar components (Lada \& Lada 2003). 
	There are also mid-infrared sources with a number density of 200 pc$^{-2}$ in the most complex regions: 
	FIR 3--5 (Nielbock et al. 2003). Several bright reflection nebulae and NIR features in this region indicate an extreme complexity.
	Twenty one of the 41 potential star-forming sites are included in this region.	
	
	$VLA~9; (PROBABLE)$-- 
	A blue-shifted outflow is detected to the northeast of VLA 9, 
	which is associated with a 24 $\mu$m source, SP 053526-050758, as shown by the dotted-line [iii] in Figure \ref{f6}$b$. 
	We consider that this blue-shifted outflow is probably driven by the 24 $\mu$m source associated with VLA 9, 
	although previous studies have reported that the blue-shifted outflow is 
	associated with MMS 10 (Williams et al. 2003, Aso et al. 2000). 
	This is because (i) MMS 10 is more likely a starless core rather than a protostellar core 
	since there is no 24 $\mu$m source associated with MMS 10, 
	and (ii) the morphology of the blue-shifted lobe seems to become more collimated 
	toward the position of the 24 $\mu$m source, but not toward MMS 10. 
	A NIR reflection nebula elongated in the same direction as the blue-shifted outflow is detected 
	to the northeast of the 24 $\mu$m source, as denoted by [13] in Figure \ref{f6}$a$, 
	which supports the idea that the blue-shifted outflow is driven by the 24 $\mu$m source. 
	Note that the red-shifted emission southwest of the 24 $\mu$m source could partially 
	be a counterpart of the blue-shifted outflow. 
	It is, however, difficult for us to conclude this, because of the contamination by another nearby outflow 
	associated with FIR 2 (see below). 
	
	$FIR~1b; (PROBABLE) $-- 
	A red-shifted CO (3--2) outflow is located southeast of FIR 1$b$ [iv] as shown in Figure \ref{f6}$b$, 
	whereas weak extended blue-shifted emission is northwest of FIR 1$b$. 
	This extended blue-shifted component may be the counterpart of the red-shifted lobe, 
	forming a bipolar outflow, as suggested by the previous CO (1--0) observations (Aso et al. 2000, Williams et al. 2003). 
	This is, however, difficult to conclude because the blue-shifted emission is 
	less localized and also is connected to the prominent outflow associated with MMS 9. 
	The red-shifted outflow is probably associated with FIR 1$b$ 
	although we cannot rule out its possible association with FIR 1$a$. 
	Note that neither FIR 1$b$ nor FIR1$a$ is associated with any 24 $\mu$m emission, 
	preventing us from identifying the driving source of the outflow. 
	There is no apparent outflow associated with FIR 1$c$.

	$FIR~2; (CLEAR)$--  There is a clear bipolar CO outflow associated with FIR 2,
	centered on the 24 $\mu$m source of SP 053524-050830, as shown in Figure \ref{f6}$b$ or \ref{f7}$b$. 
	This outflow has also been identified in CO (1--0) (Aso et al. 2000, Williams et al. 2003).
	The blue-shifted emission is located to the southwest and the red-shifted emission to the northeast of 
	24 $\mu$m source.
	
	$FIR~3; (CLEAR)$ --  
	Red- and blue-shifted CO (3--2) emissions elongated along the northeast to the southwest direction is 
	seen in Figure \ref{f7}$b$. Each of the emissions overlapping with the other has two peaks, 
	and a prominent 24 $\mu$m source, SP 053528-050935 associated with FIR 3, 
	is located between the two peaks. 
	This suggests that the outflow is driven by this 24 $\mu$m source. 
	This outflow was also identified in CO (1--0) and CO (2--1) (Aso et al. 2000, Williams et al. 2003, Wu et al. 2005).  
	Because the blue- and red-shifted lobes overlap, 
	the axis of the outflow is considered to be almost on the plane of the sky. 
	This geometrical configuration is also suggested by the position-velocity diagram cutting 
	along the outflow axis showing ``butterfly'' like velocity structure. 
	Our $JHK_s$ image presented in Figure \ref{f7}$a$ [14] shows bright knots and shell-like features to the northeast of FIR 3 
	and a reflection nebula to the southwest of FIR 3. 
	These NIR features, which are associated with the CO (3--2) outflow, also support 
	the idea that the outflow is driven by  the 24 $\mu$m source associated with FIR 3.

	In the vicinity of the peak position of FIR 3, there is another 24 $\mu$m source, 
	SP 053527-050923 (also identified as MIR 20 by Nielbock et al. 2003), located at 15$''$ northwest of FIR 3. 
	Our $JHK_s$ image presented in Figure \ref{f7}$a$ [15] shows a jet-like feature elongated to the west of this 24 $\mu$m source. 
	Interestingly, we also detected slightly extended blue-shifted CO (3--2) emission coincident 
	with the NIR jet-like feature (see Figure \ref{f7}$b$), 
	even though it is difficult for us to clearly distinguish this blue-shifted emission from the prominent 
	blue-shifted outflow driven by SP 053528-050935.
	
	$FIR~4; (MARGINAL)$--  
	The strong millimeter source FIR 4 associated with a free-free jet , VLA 12, 
	is in the near neighborhood of a 24 $\mu$m source, SP 053527-051002. 
	Although the CO (3--2) line profile toward FIR 4 shows a broad wing, 
	no clear CO outflow associated with FIR 4 was identified in our CO (3--2) map (see Figure \ref{f7}$b$). 
	Millimeter interferometric observations also showed no clear outflow associated with FIR 4 (Shimajiri et al. 2008). 
	This may be because there is strong contamination of the prominent outflow associated with FIR 3. 
	In fact, we detected a faint NIR nebula in the vicinity of FIR 4, as denoted by [16] in Figure \ref{f7}$a$ and \ref{f7}$c$, 
	which could be a hint of a potential outflow associated with FIR 4.
	
	$FIR~5; (NO)$-- A 24 $\mu$m source, SP 053527-051017 (corresponding to MIR 27 from Nielbock et al. 2003), 
	is located at 10$''$ NE of FIR 5. There is no clear bipolar CO lobe around this site. 

	$VLA~13; (CLEAR)$--
	We detected a bipolar CO (3--2) outflow along the north-south direction centered on VLA 13, 
	which coincides with a bright 24 $\mu$m source, SP 053525-051031 (see Figure \ref{f7}$b$). 
	Our $JHKs$ image shows that there is a bright cone-shaped $K_s$ reflection nebula 
	([17] in Figure \ref{f7}$a$) coincident with this outflow.  

	$FIR~6a-d$;   
	A bipolar CO (3--2) outflow categorized as ``CLEAR'' lies northeast to southwest centered on FIR 6$b$ (see Figure \ref{f8}$b$), 
	which is associated with a 24 $\mu$m source, SP 053523-051205 shown in Figure \ref{f8}$a$. 
	We also detected another bipolar CO (3--2) outflow running close to the north-south centered on FIR 6$c$. 
	This outflow was also categorized as ``CLEAR'' although its red-shifted lobe might be 
	partially the molecular ridge surrounding M 43. No 24 $\mu$m source is associated with FIR 6$c$. 
	This might be because of a very high background level due to the emission from M 43. 
	Although FIR 6$d$ is not associated with any clear CO (3--2) high velocity lobe, 
	a possible outflow we categorized as ``MARGINAL'' was found at the position of FIR 6$d$ as 
	an NIR reflection nebula denoted by [18] in Figure \ref{f8}$a$. 
	There is neither clear CO lobe nor NIR feature around FIR 6$a$. 

	$CSO~32;(MARGINAL)$--
	Although red-shifted emission is associated with CSO 32, 
	this red-shifted emission appears to be a part of the molecular ridge surrounding M 43. 

	$CSO~33;(PROBABLE)$--
	A blue-shifted CO (3--2) emission associated with CSO 33 was detected as shown in Figure \ref{f8}$b$. 
	Our $JHK_s$ image shows that a reflection nebula coincident with the blue-shifted outflow was also detected. 
	The reflection nebula extends to the east of CSO 33, which seems to be consistent with the elongation of 
	the blue-shifted emission. We cannot identify a driving source for this blue-shifted outflow since there is 
	no 24 $\mu$m source associated with CSO 33. This may be because of the high background level due to M43. 
	There is a red-shifted emission located at 15$''$ east of CSO 33, 
	and could be the counterpart of the blue-shifted outflow. 
	However, the direction of the red-shifted emission is not consistent with the extension of the reflection nebula. 
	In addition, the location of CSO 33 is not favorable for the interpretation of the red-shifted emission 
	as the counterpart of the blue-shifted outflow. 
	The red-shifted emission may be a part of the molecular ridge surrounding and probably interacting with M43.

	\subsection{\bf Notable NIR sources}
			
	We have found three notable NIR sources in the OMC-2/3 region through the observations with SIRIUS/IRSF. 
	Although these NIR sources are not associated with the 41 potential star-forming sites, 
	the $JHKs$ images we obtained show probable star-forming activities. 

	The first source is a faint reflection nebula located at $\sim$100$''$ east of MMS 2 (denoted as [2] in Figure \ref{f4}$a$). 
	This source was catalogued as 2MASS J05352454-0500214. Previous observations of H$_2$ emission at 2.2 $\mu$m 
	suggested that this reflection nebula might be a knot related to the outflow from MMS 2 (Stanke et al. 2002). 
	Our $JHK_s$ image shows extended emission with a bright point source at the center of the emission (see Figure \ref{f10}$a$).
	The extended NIR emission appears as a cone-shaped structure due to an outflow with an axis nearly pole-on. 
	These results suggest that the NIR feature probably traces an independent star-forming activity rather than a knot due to the outflow from MMS 2. 
	Strong blue-shifted and faint red-shifted CO (3--2) emissions were detected around this source (see Figure \ref{f4}$b$), 
	but their relationship with the NIR feature is not clear.

	The second source is a reddened point-like source surrounded by an extended emission elongated from northeast to southwest 
	(denoted as [a] in Figure \ref{f8}$a$). The source was catalogued as 2MASS J05352748-0511497. 
	The size of the extended emission is $\sim$8$''$ (corresponding to 3600 AU) along the major axis, while $\sim$4$''$ 
	(corresponding to 1800 AU) along the minor axis. 
	Although there is no clear CO high velocity emission associated with this source, 
	the origin of the elongated feature might be an outflow cavity.
		
	The last source is located at 80$''$ northwest of FIR 6$b$ (denoted as [b] in Figure \ref{f8}$a$). 
	The source, catalogued as 2MASS J05352307-0511490, has a cavity-like reflection nebula from nearly north to south, 
	with a silhouette disk at its middle. Although there is no clear CO (3--2) high velocity emission associated with the source, 
	the cavity-like reflection nebula is suggestive of an outflow activity. 
	
	The possible origins of these NIR sources are low-mass protostars associated with low mass molecular outflows with masses less 
	than the detection limit of our survey. 
		
\section{DATA ANALYSIS OF THE CO (3--2) EMISSION}
	Table 3 summarizes parameters toward 14 outflows categorized as CLEAR or PROBABLE. 	
	Regarding the inclination angle of the outflows of MMS 2, MMS 7, and FIR 3, we assumed 
	an inclination of 70 $^{\circ}$, since the outflow axes of these outflows  
	are nearly on the plane of the sky as seen in the outflow morphology (i.e., blue- and red-shifted lobes overlapping morphology) in the CO emission. 
	The inclination angle of the other outflows were assumed to be 45 $^{\circ}$. 
	The projected maximum size of each CO (3--2) outflow was measured at the 3$\sigma$ contour in its channel maps. 
	The velocity range of each outflow was determined using channel maps: isolated components from high velocity CO 
	emission are clearly identifiable at more than 3 $\sigma$ level within the velocity range, without significant contamination of 
	emission from the ambient component. The velocity ranges of the outflows were presented in the Table 3. 
	The maximum velocity (${\Delta}V_{\rm{max}}$) of each outflow presented in Table 3 was calculated as $|V_{\rm{sys}}-V_{\rm{max}}|$, 
	where $V_{\rm{max}}$ is either the most blue-shifted or the most red-shifted velocity of each outflow within its velocity range.
	
	Under the assumption of the Local Thermodynamical Equilibrium (LTE) condition and optically thin CO (3--2) emission, 
	we estimated the outflow mass ($M_{\rm{CO}}$) as 
	
	\begin{equation}
		M_{\rm{CO}}~[M_{\odot}]~=~ \bar{\mu}~ m_{\rm{H}}~ X[{\rm{CO}}]~\Omega_{\rm{S}}~ D^2 N_{\rm{CO}},  
	\end{equation}

	where

    \begin{equation}
		N_{\rm{CO}}~{\rm{[cm^{-2}]}}~=~ \frac{8 \pi {\nu}^3}{c^3} \times \frac{1}{(2J_{l}+3)A} 
		\times \frac{Z(T_{\rm{ex}})}{\exp(-E_{l}/kT_{\rm{ex}})[1-\exp(h{\nu}/kT_{\rm{ex}})]}
		\times \frac{{\int}T_B dV}{J(T_{\rm{ex}})-J(T_{\rm{bg}})}, 
	\end{equation}
	
	and 

	\begin{equation}
		J(T)~=~\frac{h{\nu}/k}{\exp(h{\nu}/kT)-1}.
	\end{equation}
	
	In the above expressions, $\bar{\mu}$ is the mean molecular weight of 2.33, $m_{\rm{H}}$ is the atomic hydrogen mass, 
	$h$ is Planck's constant, $k$ is Boltzmann's constant, 
	$c$ is the speed of light, $\nu$  is the line frequency, $T_{\rm{ex}}$ is the excitation temperature of the CO  transitions, 
	$\Omega$ is a solid angle of each source, $D$ is the distance to the objects, $N_{\rm{CO}}$ is the CO column density, 
	$T_{\rm{bg}}$ is the background radiation temperature, 
	$A$ is the Einstein $A$-coefficient of the transition, $Z$ is the partition function, and $E_{l}$ 
	is the rotational energy level of the lower energy state,  
	$J_{l}$ is the rotational quantum number of the lower energy state, and ${\int}T_{B}dV$ is the integrated intensity of 
	the blue- and red-shifted CO (3--2) emissions in the unit of K km s$^{-1}$. 
    Each outflow emission was integrated using the velocity range of the CO (3--2) emission presented in Table 3.
	Here, we adopted $T_{\rm{ex}}$ of 30 K which is typical peak CO (3--2) brightness temperature in the OMC-2/3 filament 
	(see CO line profiles in Figure \ref{f13}), and $X$[CO] of 10$^{-4}$ (e.g., Frerking, Langer \& Wilson 1982).	
	The detection limit of the mass corresponds to as small as 10$^{-4}~M_{\odot}$ within a single beam 
	\footnote{We derived the detection limit of the outflow mass assuming that the 3 $\sigma$ emission level and $\delta$V=1.08 km s$^{-1}$}.
	The size and velocity of the outflow after the correction for the projection with the 
	inclination angle, $i$, are estimated to be $R~=~R_{\rm{max}}/\sin i$, and $V~=~{\Delta}V_{\rm{max}}/\cos i$, respectively.
	Here, a pole-on outflow was defined as an inclination angle of 0$^{\circ}$.
	We obtained the outflow dynamical time, $t_d~ = ~R/V$, 
	mass outflow rate, $\dot{M}_{\rm{CO}}~=~M_{\rm{CO}}/t_d$, 
	momentum, $P_{\rm{CO}}~=~M_{\rm{CO}}{\times}V_{\rm{\rm{max}}}$, 
	and momentum flux, $F_{\rm{CO}}~=~P_{\rm{CO}}/t_d$. 

\section{PROPERTIES OF THE PROTOSTELLAR CORES IN OMC-2/3}
	In Table 4, we list the derived physical properties of the individual protostellar cores having CO outflows categorized as CLEAR. 
	These individual sites have compact infrared sources (IRAS sources and/or Spitzer sources), which are considered to be 
	the driving sources of the associated CO outflows. The bolometric luminosity of each driving source was estimated from the total 
	flux density integrated from 8 $\mu$m to 1.3 mm continuum data. Since the IRAS sources in these regions have only upper limits at 
	100 $\mu$m, upper and lower limits of the bolometric luminosity for each driving source were derived with and without the flux density 
	at 100 $\mu$m, respectively. The mass of the envelope surrounding each of the driving sources was estimated from the 1.3 mm 
	flux density observed by Chini et al. (1997) and Nielbock et al. (2003). On the assumption that the 1.3 mm continuum is optically thin, 
	the mass is derived using the following formula;
		
		\begin{equation}
			M_{\rm{dust}} = \frac{S_{\lambda}d^2}{{\kappa}_{\lambda}B_{\lambda}(T_{\rm{dust}})}
		\end{equation}
	where, $\kappa_{\lambda}$ is the mass-absorption coefficient of the dust grains, 
	$B_{\lambda}$ is the Planck function, $d$ is the distance to the OMC-2/3 region (450 pc;Genzel \& Stutzki 1989), 
	and $S_{\lambda}$ is the total flux density of the continuum emission within the 11$''$ radius 
	(corresponding to 5000 AU) as repeated by Chini et al. 1997. 
	We adopted a dust opacity of $\kappa_{\nu}=0.037$ cm$^{-2}$ g$^{-1} 
	(400~{\mu}m/\lambda)^{\beta}$ (Ohashi et al. 1996), $\beta$=2, and $T_{\rm{dust}}$=20 K (i.e., typical temperature of these 
	cores from Cesaroni et al. 1994, Chini et al. 1997).  
	The estimated envelope masses are found to be $\leq$1.0 to 7.7 $M_{\odot}$. 	
	
	What is the evolutionary status of these sources? The spectral slope between near- and mid-infrared is often used to identify 
	the evolutionary status of young stellar objects (e.g., Adams et al. 1988). We used the 2 $\mu$m and 10 $\mu$m data 
	taken by Nielbock et al. (2003) to derive spectral slopes of our 10 cores: 6 sources (MMS 2, MMS 7, FIR 2, FIR 3, VLA 13, and FIR 6$b$) 
	were found to show positive slopes (e.g., rising spectrum toward 10 $\mu$m), suggesting that these sources are Class I type of 
	objects. Four out  of the 10 sources (SIMBA $a$, MMS 5, MMS 9, and FIR 6$c$), on the other hand, 
	were not detected at 10 $\mu$m  and shorter wavelengths with the ground-based telescopes. 
	Since these 4 sources are as luminous as the 6 detected sources mentioned above (see Table 4), 
	they should not be too faint to be detected at 10 $\mu$m and shorter wavelengths. 
	A possible reason for this non-detection at near-infrared is that they are even more deeply embedded than the 
	6 detected sources. If this is the case, these 4 non-detected sources would be even younger than the 6 detected sources, 
	i.e., Class 0 type of objects. Hence, these 4 sources are categorized as Class 0 type of objects in our paper. 
	We note that the ratio of submillimeter luminosity  ($L_{\rm{sbmm}}$) to bolometric luminosity ($L_{\rm{bol}}$ ) 
	is often used to distinguish Class 0 sources from Class I sources (Andre et al. 1993). 
	It is, however, not easy for us to take this approach for our sample sources because of the large uncertainty 
	of the bolometric luminosities we estimated, as described above.

\section{DISCUSSION}
\subsection{Comparison with previous CO (1--0) observations}
	In addition to our CO (3--2) observations, outflow observations in the OMC-2/3 region were also conducted by Aso et al. (2000) 
	with the Nobeyama 45 m telescope, and Williams et al. (2003) with the FCRAO 14 m telescope and the BIMA array. 
	These previous observations were done in CO (1--0). In this section, we compare our results with those obtained 
	in the previous observations.

	Out of 14 outflows (10 are categorized as CLEAR and 4 are categorized as PROBABLE) we identified, 
	the one associated with CSO 33 located in the most southern part of the OMC-2/3 region was not included in the previous outflow observations. 
	Out of the rest, eight outflows associated with MMS 2, MMS 5, MMS 7, MMS 9, VLA 9, FIR 1$b$, FIR 2 and FIR 3 were also detected in 
	both of the previous observations although one of them, associated with VLA 9, was considered to be associated with MMS 10 
	in the previous observations (see section 3.3.3; see also below). The physical parameters of these eight outflows derived from 
	our CO (3--2) data (see Table 4) are roughly in agreement with those derived from the previous CO (1--0) data. 
	On the other hand, five outflows associated with SIMBA $a$, SIMBA $c$, VLA 13, FIR 6$b$, and FIR 6$c$ have been detected only in our CO (3--2) observations.

	Why were these five outflows not detected in the previous CO (1--0) observations? 
	One reason is the complexity of the region; some parts of the OMC-2 region form young stars as groups, 
	and as a result, it is difficult for us to identify each outflow. 
	For example, even though previous observations detected 
	high velocity red-shifted CO (1--0) emission near VLA 13, it was not identified as an outflow associated with VLA 13. 
	In this study, it was identified as an outflow associated with VLA 13 because of the near- and mid-IR data as well as the CO (3--2) data. 
	A similar case is the outflow associated with VLA 9; previous studies misidentified this outflow as that associated with MMS 10 
	because of the lack of near- and mid-infrared data.

	Another possibility is that high velocity emission is weaker in the CO (1--0). 
	In the case of outflows associated with SIMBA $a$ and SIMBA $c$, no high velocity emission was clearly detected 
	in CO (1--0) in the previous observations. These outflows are weak even in the CO (3--2) emission. 
	Similarly, outflows associated with FIR6 $b$ and FIR6 $c$ are significantly 
	weak in CO (1--0) even though they are quite strong in CO (3--2). 

	Are outflow emissions systematically weaker in CO (1--0) as compared with CO (3--2)? 
	In order to examine this, we compared the intensity of CO (1--0) and (3--2) emissions toward the outflow lobes. 
	For this comparison, we used new CO (1--0) outflow maps of the OMC-2/3 region, recently taken using 
	the Nobeyama 45 m telescope with the OTF mode (Kawabe et al.) in addition to our CO (3--2) maps taken by ASTE. 
	Since the CO (1--0) maps have a higher angular resolution (22$''$) than the CO (3--2) maps (26$''$), 
	the CO (1--0) maps were convolved with a beam of 26$''$.  
	For the outflows significantly detected in both CO (1--0) and (3--2), CO (3--2)/CO (1--0) intensity ratios were estimated to be 1-3, 
	suggesting that CO (3--2) emission is at least comparable to, or brighter than CO (1--0) for the outflows in the OMC-2/3 region. 

	It is interesting to consider what kind of physical conditions could make the CO (3--2) emission a few times stronger than the CO (1--0) emission. 
	With the Large Velocity Gradient (LVG) calculations, the CO (3--2)/CO(1--0) ratio would be more than unity when 
	H$_2$ density range of 1.5$\times$10$^3$ to 2$\times$10$^4$ cm$^{-3}$ and the kinematic temperature of 40 K. 
	The gas kinetic temperature at some of outflow lobes actually estimated to be more than 40 K from CO (3-2) line profiles. 
	Hence, these molecular outflows we observed in the OMC-2/3 region probably have high density gas of $\sim$10$^4$ cm$^{-3}$. 
	Such high density and high temperature gas were found in the molecular outflow associated with L1157, 
	one of the best-studied molecular outflows (Hirano \& Taniguchi 2001).

\subsection{Outflow Properties: OMC-2/3 Region versus Other Regions}

	In the following section, we compare the physical nature of the outflows in the OMC-2/3 region with 
	those found through previous systematic outflow searches in other star-forming regions. 
	In order for us to minimize the ambiguity of the outflow identification, 
	we limited our outflow sample in the OMC-2/3 region to those categorized as CLEAR or PROBABLE 
	in the sections 6.2.1 and those categorized as CLEAR in the section 6.2.2.
	
	\subsubsection{Detection Rate of the outflows}
	Among the 41 potential star-forming sites listed in Table 1, 
	14 sites were found to be associated with CO outflows categorized as CLEAR or PROBABLE. 
	When we simply calculate the CO outflow detection rate using this result, the rate is 34 \%. 
	On the other hand, previous outflow searches in low-mass star-forming regions (e.g., Bontemps et al. 1996, Hogerheijde et al. 1998) 
	and in high-mass star-forming regions (e.g., Shepherd \& Churchwell 1996, Zhang et al. 2001) showed that the detection rate is more than 80 \%. 
	This might suggest that the CO outflow detection rate in the OMC-2/3 region is significantly lower than in the other star-forming regions. 
	We should note, however, that 23 of the 41 potential star-forming sites (49 \%) do not have Spitzer 24 $\mu$m source; 
	these sites are the so-called prestellar cores, where no infrared embedded sources have been found. 
	In contrast, the previous outflow searches mentioned above were biased toward IRAS sources. 
	When we limit our regions to those with 24 $\mu$m sources, then the CO outflow detection rate becomes 67 \%. 
	This is comparable to, but not as high as those of the previous searches. 
	This is probably because; (i) a part of the ambient molecular cloud in the OMC-2/3 region shows CO emissions with higher velocities, 
	which are difficult to distinguish from molecular outflows; (ii) some of the 24 $\mu$m sources seem to be more evolved with less surrounding material.

	\subsubsection{Relation between outflow parameters and driving source properties}
	Previous studies have discussed the driving mechanisms and the physical properties of CO outflows 
	(e.g., Lada 1985, Cabrit et al. 1992, Bontemps et al. 1996). 
	Their main conclusions are as follows: (i) CO molecular outflows driven by a highly collimated flow (i.e., primary jet) are observed as entrained surrounding 
	molecular gas, (ii) Energy of each CO outflow is mainly determined by the luminosity of the central driving source, which is probably related to the 
	processes of mass accretion onto the central stars. 
	In this subsection, we investigate the relationship between the physical parameters of the molecular outflows and of their driving sources.

	In Figure \ref{f11}, we plot the momentum flux of the outflow ($F_{\rm{CO}}$) as a function of the bolometric luminosity of 
	the outflow driving source ($L_{\rm{bol}}$) for 10 sources from our sample. 
	For comparison, Figure \ref{f11} also shows the data of 101 CO outflows associated with young stellar objects (YSOs) in low- and  
	high-mass star-forming regions, which were previously observed in CO with single-dish telescopes (Bontemps et al. 1996 and Hogerheijde et al. 
	1998 for low-mass star-forming regions, Beuther et al. 2002, Zhang et al. 2005 for high-mass star-forming regions). 
	
	Figure \ref{f11} clearly shows a positive correlation between $F_{\rm{CO}}$ and $L_{\rm{bol}}$. 
	The correlation is valid in a wide range of $L_{\rm{bol}}$ (0.1 ${\leq}L_{\rm{bol}}{\leq}~10^5{L_{\odot}}$) and 
	$F_{\rm{CO}}$ ($10^{-6}{\leq}{F_{\rm{CO}}}{\leq}0.1~M_{\odot}$ km s$^{-1}$ yr$^{-1}$). 
	Sources in low-mass star-forming regions have lower $F_{\rm{CO}}$ and $L_{\rm{bol}}$, 
	while those in high-mass star-forming regions have higher $F_{\rm{CO}}$ and $L_{\rm{bol}}$. 
	This conclusion has already been reported by Beuther et al. (2002). 
	The sources in the OMC 2/3 region are located in this plot between the sources in low- and high-mass star-forming 
	regions, and bridge a gap between them. The sources in the OMC-2/3 region have intermediate 
	$F_{\rm{CO}}$ and $L_{\rm{bol}}$ in Figure \ref{f11}, which naturally suggests that they are most probably intermediate-mass YSOs. 
		
	It is interesting to consider why both $L_{\rm{bol}}$ and $F_{\rm{CO}}$ can be good indicators of the mass of the YSOs. 
	For low- and intermediate-mass YSOs, their bolometric luminosities are most probably dominated by the accretion luminosity 
	($L_{\rm{acc}}{\equiv}GM_{\ast}{\dot{M}_{\rm{acc}}}/R_{\ast}$). 
	If we assume that the stellar radius ($R_{\ast}$) does not significantly change during protostellar evolution, 
	the relation suggests that a higher $L_{\rm{bol}}$ indicates either 
	a higher mass accretion rate or a higher central stellar mass, or both. 
	Since a higher mass accretion rate results in a higher stellar mass as long as the mass accretion time scale is similar 
	among low- and intermediate- mass protostars, it is fair to assume that $L_{\rm{bol}}$ is an 
	indicator for the mass of YSO for low- and intermediate-mass YSOs. For high-mass YSOs, on the other hand, it is not always true that $L_{\rm{bol}}$ 
	is dominated by the accretion luminosity since some of them reach the main sequence when they are still embedded in molecular clouds. 
	If this is the case, $L_{\rm{bol}}$ is basically the same as the stellar luminosity, indicating that $L_{\rm{bol}}$ is an indicator of the stellar mass. 

	How about $F_{\rm{CO}}$? It is not very clear how $F_{\rm{CO}}$ can be an indicator of the mass of YSOs. Bontemps et al. (1996) suggested that 
	$F_{\rm{CO}}$ has a linear relationship with the rate of mass accretion onto the central star for low-mass YSOs. 
	This is because the most plausible energy source for a jet/wind, which is considered to drive a CO outflow, 
	is the gravitational energy released by infall and/or accretion onto  
	the central protostar. If this is valid for low- to high-mass YSOs, 
	$F_{\rm{CO}}$ is indirectly related to the central stellar mass which is more or less determined by the mass accretion rate.
	
	In Figure \ref{f12}, we plot $F_{\rm{co}}$ as a function of the envelope mass ($M_{\rm{env}}$). 
	In the same way as in Figure \ref{f11}, we plot sources in the low- and high-mass star-forming regions (Bontemps et al. 1996, 
	Hogerheijde et al. 1998, and Beuther et al. 2002) as well as those in the OMC-2/3 region. 
	Figure \ref{f12} clearly shows a good correlation between $F_{\rm{co}}$ and $M_{\rm{env}}$. 
	Note that Bontemps et al. (1996) also showed the same correlation 
	for low-mass YSOs, whereas Figure \ref{f12} includes not only low-mass Class 0 and Class I sources, but also intermediate- and high-mass YSOs. 
	Figure \ref{f12} shows that more massive YSOs are associated with higher envelope masses, suggesting that 
	$M_{\rm{env}}$ is a good indicator of the YSO mass. 
	This is probably not surprising because the core mass function often shows a relation similar to the initial mass 
	function in various star-forming regions (e.g., Testi \& Sargent, 1998, Motte et al. 1998, Nutter \& Ward-Thompson 2007). 

	As discussed above, our sample sources are considered to be intermediate-mass YSOs. 
	In the OMC2/3 region, however, low-mass YSOs should form as well (see Section 3.4). 
	Why are no low-mass YSOs included in our sample sources? One probable reason is that our sample sources are 
	selected based on 1.3 mm, 350 $\mu$m, and 3.6 cm observations, as described in Section 3.1. 
	The mass sensitivity (3 $\sigma$) of the dusty envelope detected in the 1.3 mm observations is roughly 1 $M_{\odot}$. 
	For comparison, a typical value of  the envelope mass associated with low-mass protostars are 0.01 to 1 $M_{\odot}$ 
	(e.g., Bontemps et al. 1996, Hogerheijde et al. 1998; see Figure 12). 
	Therefore, most low-mass candidates may have been missed in our sample sources. 
	In other words, our observations were biased toward the intermediate-mass proto-stellar cores.
	
	In addition to the discussion whether the outflow momentum flux depends on the central protostellar mass, 
	it is also important to consider the evolutional effect on the outflow momentum flux. Bontemps et al. (1996) 
	found a decline in the CO momentum flux from the Class 0 to Class I phase, suggesting that the outflow power declines with age. 
	This result would be related to the decline of the mass accretion rate during the protostellar evolution. 
	Our observational results were also examined from this point of view. In Figures 11 and 12, Class I sources in our sample 
	were plotted as open diamonds while Class 0-like sources in our sample were plotted as filled diamonds. 
	Figures 11 and 12 show no clear difference in the distribution of the momentum flux between Class 0 and Class I source in our samples. 
	Why was no evolutionary trend seen in our samples? One possible reason is that the number of our sample sources may be 
	too small to see such an evolutionary trend. 

	We should, however, note that there might be one example that shows a possible evolutionary trend.  
	The source is SIMBA $a$, one of the Class 0-like sources in our sample. 
	This source has particularly small momentum flux in comparison with the other sources in our sample, and seems not to follow 
	the $M_{\rm{env}}-F_{\rm{CO}}$ correlation in Figure 12. Even though $M_{\rm{env}}$ of SIMBA $a$ is as massive as those of 
	our other samples, its $F_{\rm{CO}}, 3.6 {\times}10^{-6}~M_{\odot}~km~s^{-1}~{yr^{-1}}$, is 1-2 orders of magnitude smaller than others. 
	This source might be in the earliest stage of the protostellar evolution, in which the mass of the envelope is large 
	whereas the accretion/outflow activity is still less.

\section{SUMMARY}

We have carried out an extensive outflow survey in the submillimeter CO (3--2) emission and at $JHK_s$ -bands 
toward the Orion Molecular Cloud-2 and -3 region which includes 41 potential star-forming sites. 
Main results are summarized as follows: 

\begin{itemize}

	\item	From CO (3--2) and $JHK_s$ observations in addition to the multiwavelength continuum data, 
	we detected 14 secure outflows, a half of which are newly identified in the OMC-2/3 region. 
	We also found Spitzer 24 $\mu$m sources toward 12 out of the 14 outflows as their 
	driving source.  In addition to these CO outflows, seven sources having NIR features suggestive of outflows were identified.  
	This high fraction of newly identified CO outflows suggests that CO (3--2) emission may be a better outflow tracer than CO (1--0).
	
	\item	The outflow detection rate in OMC-2/3 is 34 \% which is significantly lower than those in previous low- and 
	high-mass studies (i.e., $\gtrsim$ 80\%). This is because the 41 potential star-forming sites probably include 
	23 potential prestellar cores. When we limit our sample to those with 24 $\mu$m sources,
	the CO outflow detection rate becomes 67\% which is comparable to the previous low- and high- mass outflow studies.
	
	\item	Physical properties of these outflows and their possible driving sources were derived. 
	Our results bridge a gap between previous low- and high-mass outflows, and the derived parameters 
	show $L_{\rm{bol}}-F_{\rm{CO}}$ and $M_{\rm{env}}-F_{\rm{CO}}$ tight correlations regardless 
	of the mass of the driving source. These results naturally suggest that the detected outflows 
	in the OMC-2/3 region most probably are driven by intermediate-mass protostars. 
	
\end{itemize}
\vspace{2.0cm}
 
\acknowledgments
The authors are grateful to T. Sawada, and K. Nakanishi for their support during the ASTE observations.
We acknowledge A. Ishihara and SIRIUS /IRSF team for supporting the SIRIUS/ IRSF observations. 
We also acknowledge P.T.P. Ho, Hsieh P., Y-N. Su, K. Saigo, J. Karr, and M. Puravankara for fruitful comments. 
This publication used archival data from the Spitzer Space Telescope.
We thank D. Johnstone and D. Lis for providing us the submillimeter continuum data taken by JCMT and CSO, respectively.
We also thank the referee for the constructive comments that have helped to improve this manuscript.
A part of this study was financially supported by the MEXT Grant-in-Aid for Scientific Research on Priority Areas No. 15071202 and No. 16077204.
S. Takahashi was financially supported by the Japan Society for the Promotion of Science (JSPS) for Young Scientists.

\clearpage

\begin{figure}
\epsscale{0.70}
\plotone{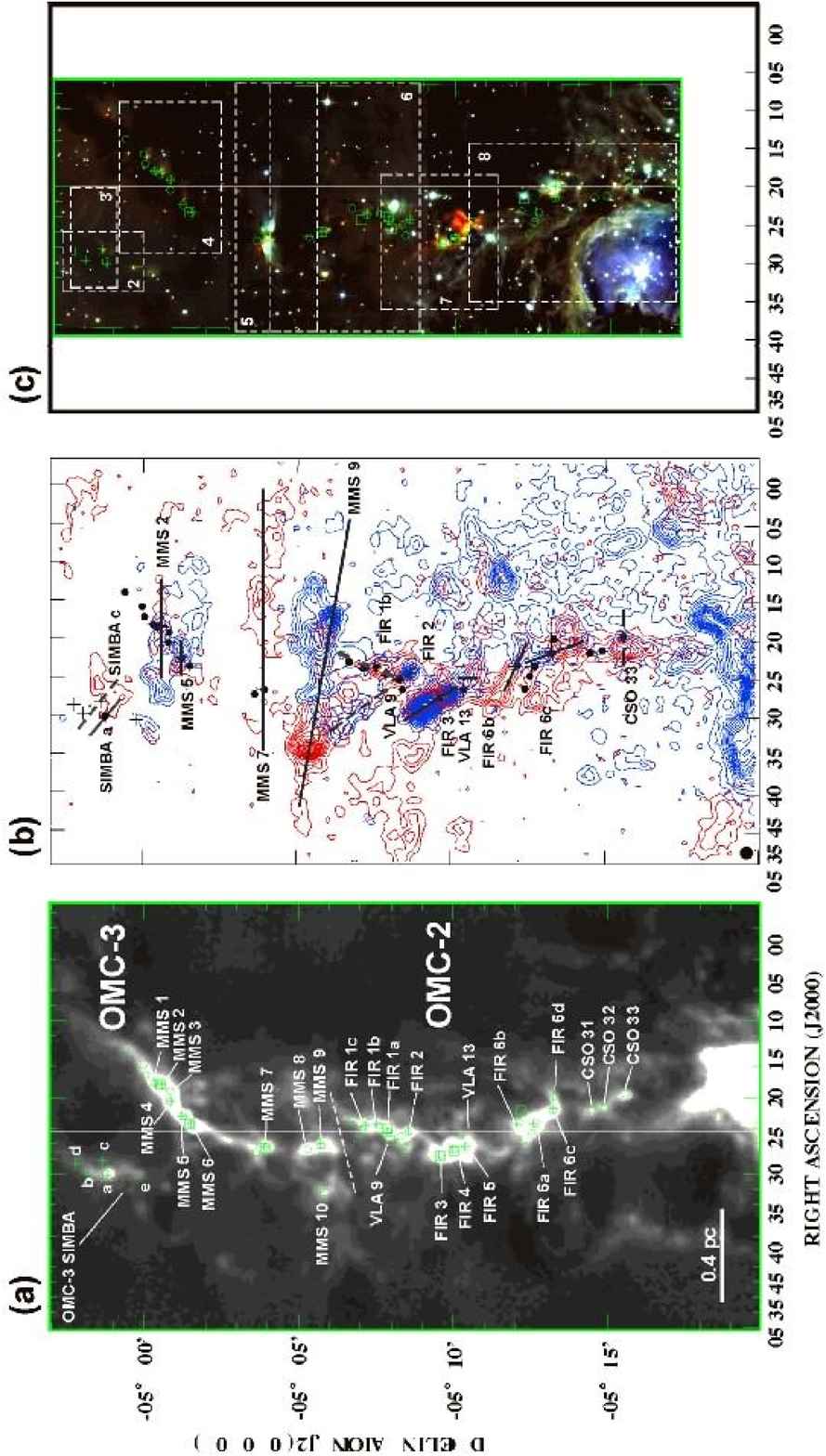}
\caption{\scriptsize $(a)$ 850 $\mu$m dust continuum map of the OMC-2/3 region from Johnstone et al. (1999).  
$(b)$ Averaged intensity map of the CO(3--2) emission taken with the ASTE with a velocity range of 
blue and red contours are $V_{\rm{LSR}}$ = -6.2 to 7.8 km s$^{-1}$ and $V_{\rm{LSR}}$ = 14.3 to 27.3 km s$^{-1}$, respectively. 
Contour intervals are 0.45 K km s$^{-1}$ starting at 0.45 K km s$^{-1}$. 
Solid and dashed lines show outflow axes categorized as CLEAR (SIMBA $a$, MMS 2, MMS 5, MMS 7, MMS 9, FIR 2, FIR 3, VLA 13, FIR 6$b$, and FIR 6$c$) 
and PROBABLE (SIMBA $c$,VLA 9, FIR1$b$, and CSO 33), respectively.
$(c)$ The $JHK_s$ composite image taken by the IRSF/SIRIUS. Pseudo colors of blue, green, and red show the data 
of $J$-, $H$-, and $K_s$- bands, respectively. Dashed squares in the Figure $c$ show the focusing areas in the Figure \ref{f2} to \ref{f8}.  
Numbers in the squares correspond to the figure numbers.
Crosses, dots, and squares indicate positions of 1.3 mm, 350 $\mu$m, and 3.6 cm sources, respectively
(from Chini et al. 1997, Lis et al. 1998, and Nielbock et al. 2003)
\label{f1}}
\end{figure}

\begin{figure}
\epsscale{1.00}
\plotone{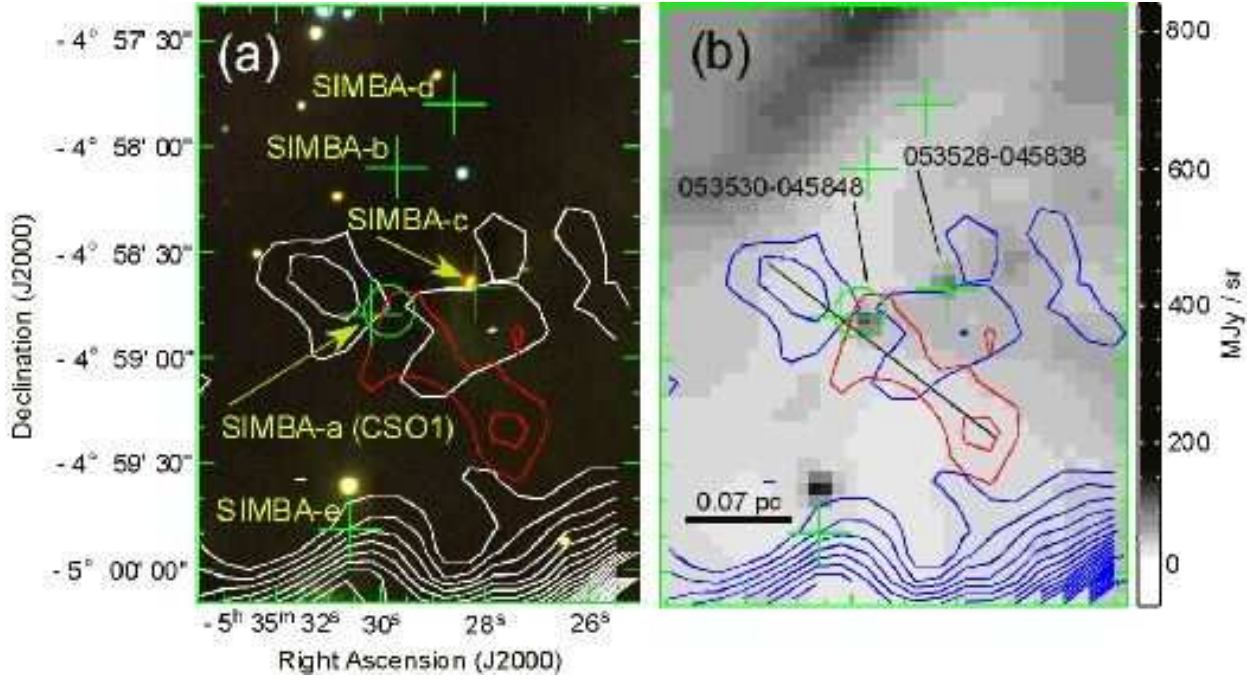}
\caption{
  Images of SIMBA $a$; 
  (a) $JHK_s$ image taken by the IRSF/SIRIUS (color) superposed on 
  the CO (3--2) emission taken with the ASTE telescope (contours). 
  (b) 24 $\mu$m image taken with Spitzer/MIPS  (grey scale) superposed 
  on the CO (3--2) emission taken with the ASTE telescope (contours).
  The blue (white in panel $a$) and red contours show the blue- and red-shifted components in the velocity range of 
  7.8 km s$^{-1}~{\leq}~V_{\rm{LSR}}~{\leq}$  8.9 km s$^{-1}$ and 15.4 km s$^{-1}~{\leq}~V_{\rm{LSR}}~{\leq}~$  16.5 km s$^{-1}$, respectively. 
  Contour intervals are  1.13 K km s$^{-1}$ starting at  2.25 K km s$^{-1}$.
  A solid line in panel $b$ in Figure $a$ indicates the outflow axis categorized as CLEAR. 
  Crosses and open circles indicate positions of 1.3 mm and 350 $\mu$m sources, respectively (from Nielbock et al. 2003, Lis et al. 1998). 
  \label{f2}}
\end{figure}

\begin{figure}
\epsscale{1.00}
\plotone{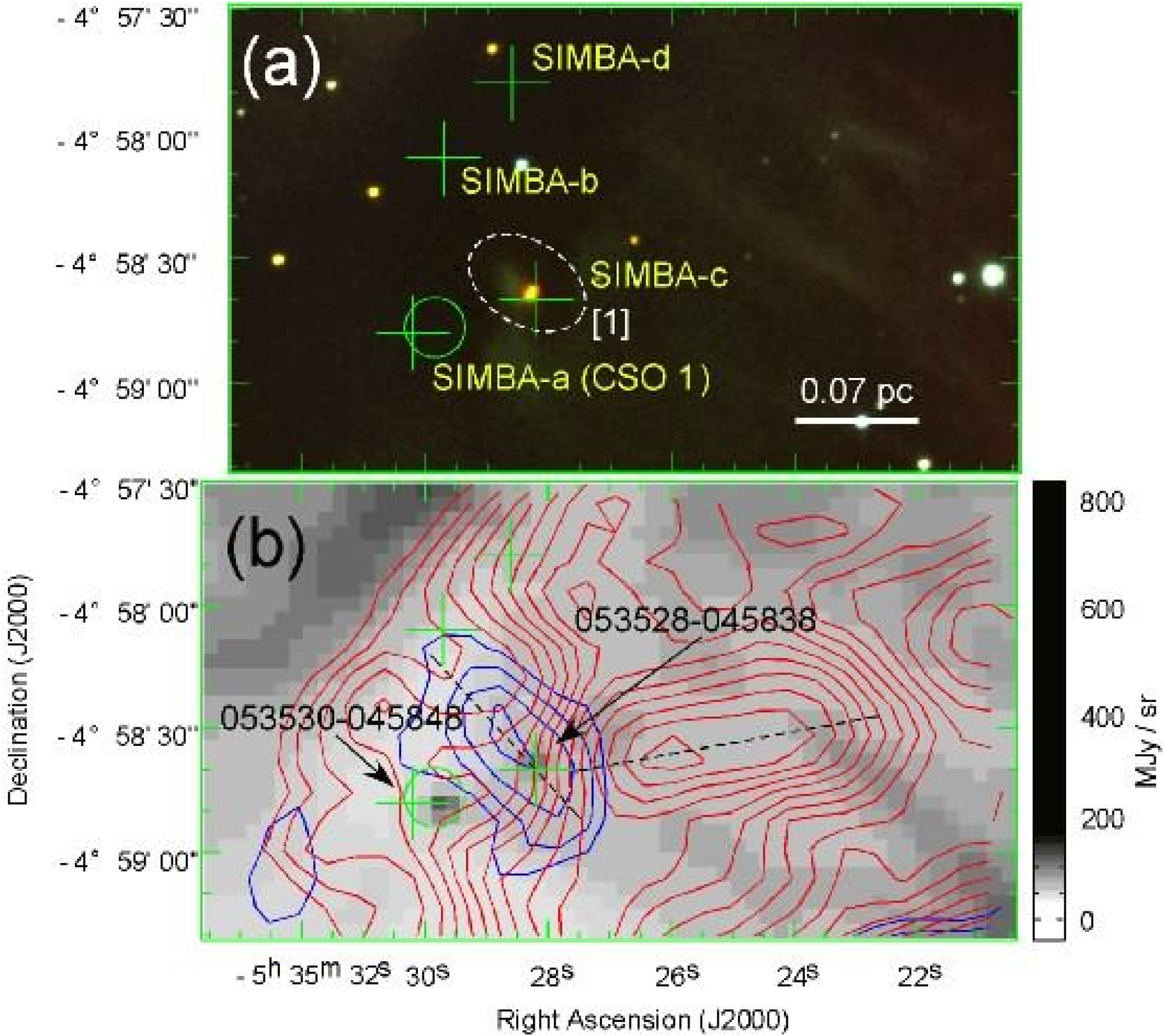}
\caption{
  Images  of SIMBA $c$; 
  (a) $JHK_s$ image taken by the IRSF/SIRIUS. 
  (b) 24 $\mu$m image taken with Spitzer/MIPS (grey-scale) superposed 
  on the CO (3--2) emission taken with the ASTE telescope (contours).
  The blue and red contours in panel $b$
  show the blue- and red-shifted components in the velocity range of 
  8.9 km s$^{-1}~{\leq}~V_{\rm{LSR}}~{\leq}~$  10.1 km s$^{-1}$ 
  and 13.5 km s$^{-1}~{\leq}~V_{\rm{LSR}}~{\leq}~$  14.2 km s$^{-1}$, respectively. 
  Contour intervals are  1.50 K km s$^{-1}$ starting at  10.0 K km s$^{-1}$.
  A dashed line in panel $b$ indicates direction of the outflow categorized as PROBABLE.
  Crosses and open circles indicate positions of 1.3 mm and 350 $\mu$m sources, respectively (from Nielbock et al. 2003, Lis et al. 1998). 
  Dashed ellipse shows the position of NIR feature presented in Table 2.
  \label{f3}}
\end{figure}

\begin{figure}
\epsscale{1.00}
\plotone{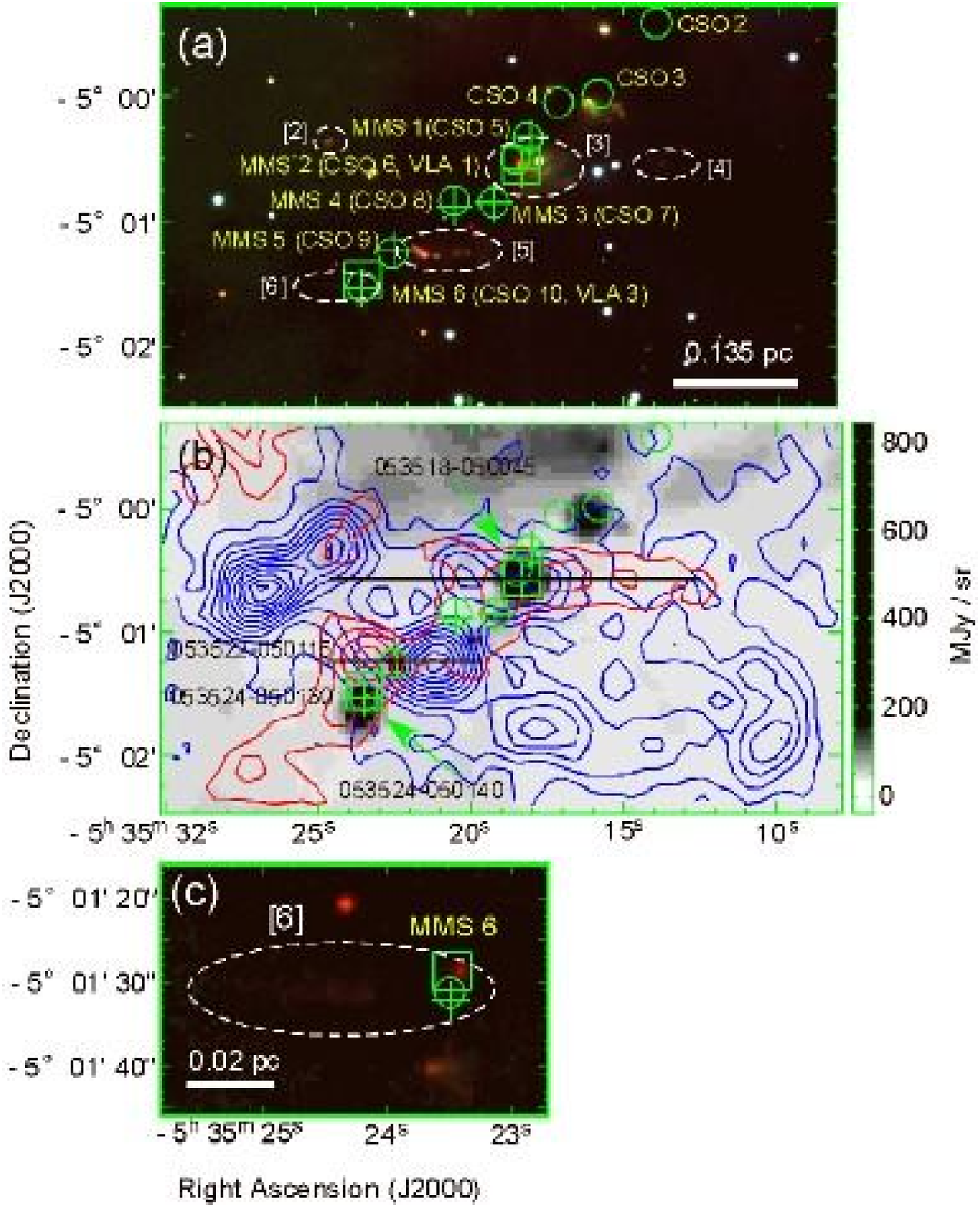}
\caption{\scriptsize 
   Images in the northern part of the OMC-3 region.  
  (a) $JHK_s$ image taken by the IRSF/SIRIUS. 
  (b) 24 $\mu$m image taken with the Spitzer/MIPS (grey-scale) superposed on 
  the CO (3--2) high-velocity emission taken with the ASTE telescope (contours).
  The blue and red contours show the blue- and red-shifted components in the velocity range of 
  1.3 km s$^{-1} \leq V_{\rm{LSR}}~ \leq$  6.7 km$^{-1}$ and 14.3 km s$^{-1}~\leq~V_{\rm{LSR}}~\leq $  15.4 km$^{-1}$, respectively. 
  Contour intervals are 2.4 K km s$^{-1}$ starting at  3.6 K km s$^{-1}$ and 1.46 K km s$^{-1}$ starting at  2.19 K km s$^{-1}$ , respectively.
  (c) Zoom up view of the $JHK_s$ image toward MMS 6. 
  A solid line in panel $b$ indicates the outflows categorized as CLEAR associated with MMS 2 and MMS 5.
  Crosses, open circles, and open squares indicate positions of 1.3 mm, 350 $\mu$m and 3.6 cm sources, respectively 
  (from Chini et al. 1977, Lis et al. 1998, Reipurth et al. 1999). 
  Dashed ellipses show the positions of NIR feature presented in Table 2.
  \label{f4}}
\end{figure}

\begin{figure}
\epsscale{0.5}
\plotone{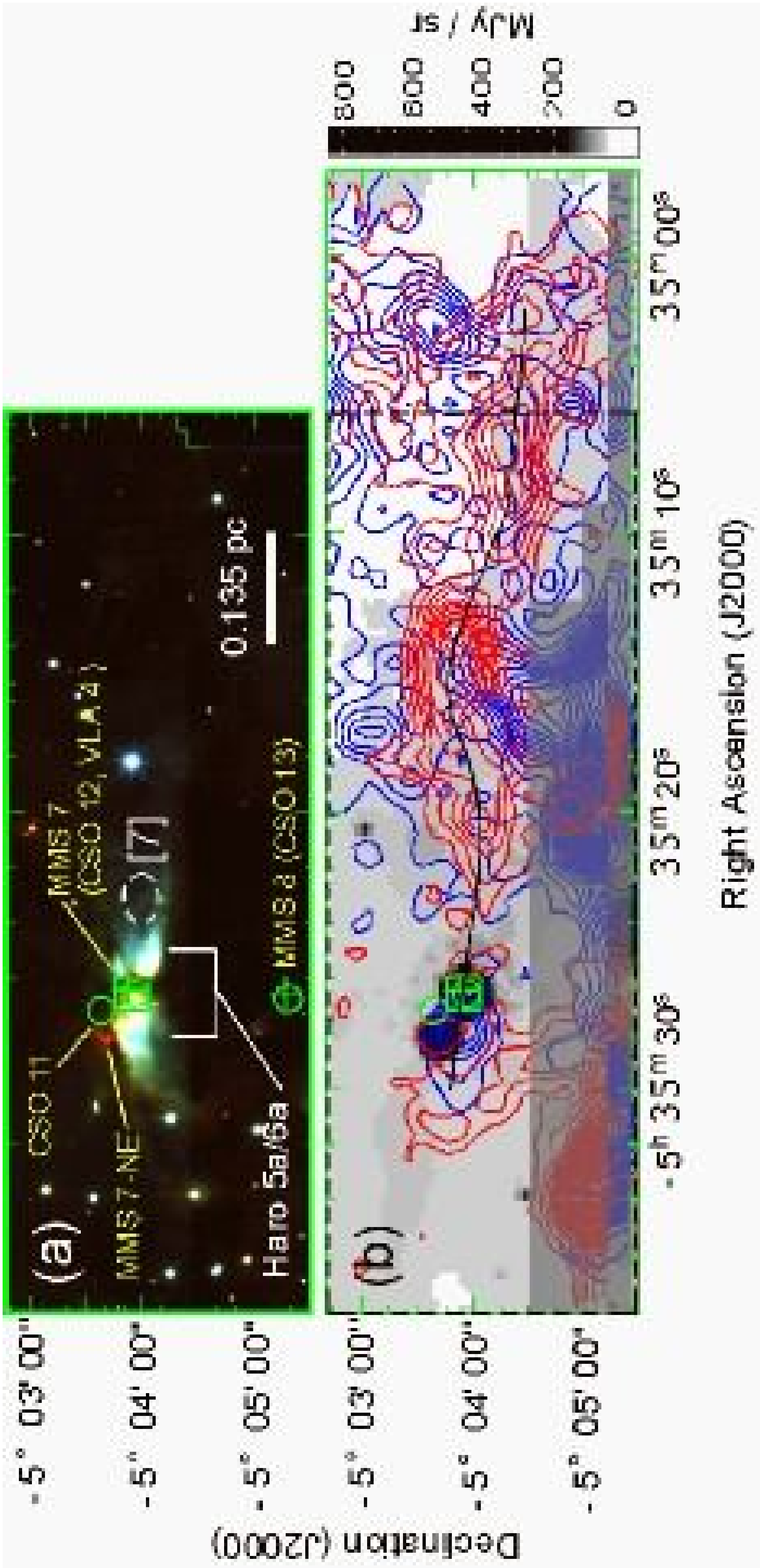}
\caption{
  Images of MMS 7. 
  (a) $JHK_s$ image taken by the IRSF/SIRIUS. 
  (b) 24 $\mu$m image taken with Spitzer/MIPS (grey-scale) superposed on 
  the CO (3--2) emission taken with the ASTE telescope (contours).
  The blue and red contours in panel $b$ show the blue- and red-shifted components in the velocity range of 
  6.7 km s$^{-1} \leq V_{\rm{LSR}}~ \leq$  7.8 km$^{-1}$ and 14.3 km s$^{-1}~\leq~V_{\rm{LSR}}~\leq $  16.5 km$^{-1}$, respectively. 
  Contour intervals are 1.26 K km s$^{-1}$ starting at 
  1.89 K km s$^{-1}$ and 1.74 K km s$^{-1}$ starting at  2.61 K km s$^{-1}$, respectively.
  A solid line in panel $b$ traces the local peaks of the CO outflow associated with MMS 7 which are categorized as CLEAR.
  Crosses, open circles, and open squares indicate positions of 1.3 mm, 350 $\mu$m and 3.6 cm sources, respectively 
  (from Chini et al. 1977, Lis et al. 1998, Reipurth et al. 1999). 
    Dashed ellipse shows the position of NIR feature presented in Table 2.
  ``Grey zone '' in the bottom part of Figure $b$ show the bipolar outflow from the source,MMS 9, located south of MMS 7.
  \label{f5}}
\end{figure}
\newpage

\begin{figure}
\epsscale{1.00}
\plotone{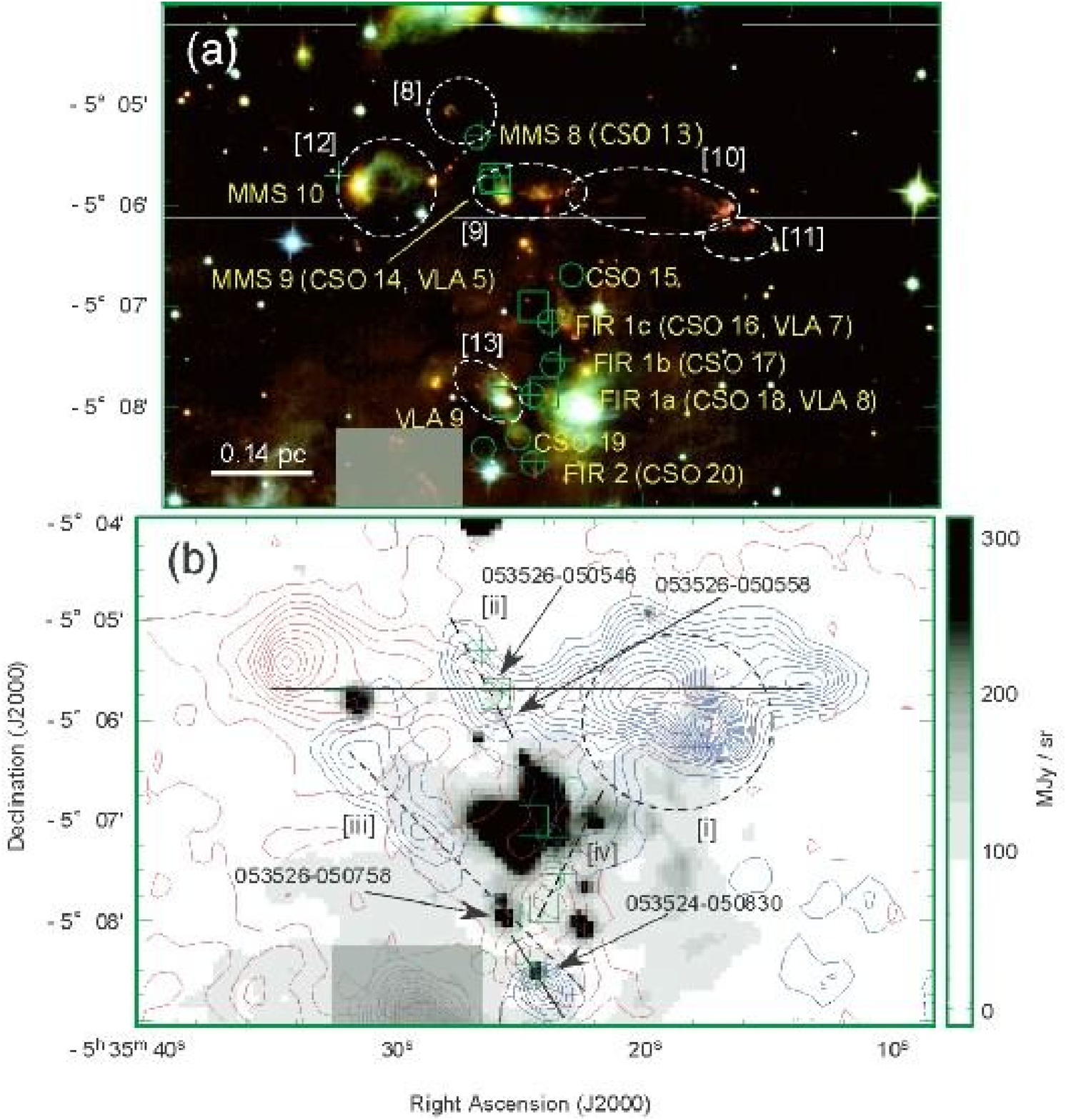}
\caption{
   Images in the MMS 8 to FIR 1 region. 
  (a) $JHK_s$ image taken by the IRSF/SIRIUS. 
  (b) 24 $\mu$m image taken with the Spitzer/MIPS (grey-scale) superposed on 
  the CO (3-2) emission taken with the ASTE telescope (contours).
  The blue and red contours show the blue- and red-shifted components in the velocity range of 
  -1.9 km s$^{-1} \leq V_{\rm{LSR}}~ \leq$  6.7 km$^{-1}$ and 14.3 km s$^{-1}~\leq~V_{\rm{LSR}}~\leq $  21.9 km$^{-1}$, respectively. 
  Contour intervals are 6.90 K km s$^{-1}$ starting at  4.14 K km s$^{-1}$ and 4.62 K km s$^{-1}$ starting at  7.70 K km s$^{-1}$, respectively. 
  Solid-, dashed- and dash-dotted- lines in panel $b$ show the outflows categorized as CLEAR (EW outflow associated with MMS 9),
  PROBABLE (associated with VLA 9 and FIR 1$b$), MARGINAL (NE-SW outflow associated with MMS 9), respectively. 
  Crosses, open circles, and open squares indicate positions of 1.3 mm, 350 $\mu$m and 3.6 cm sources, respectively 
  (from Chini et al. 1977, Lis et al. 1998, Reipurth et al. 1999). 
  Hatched areas are outflow contamination from southern part, FIR3 (see Figure \ref{f7}).
  Dashed ellipses show the positions of NIR feature presented in Table 2.
\label{f6}}
\end{figure}

\begin{figure}
\epsscale{1.00}
\plotone{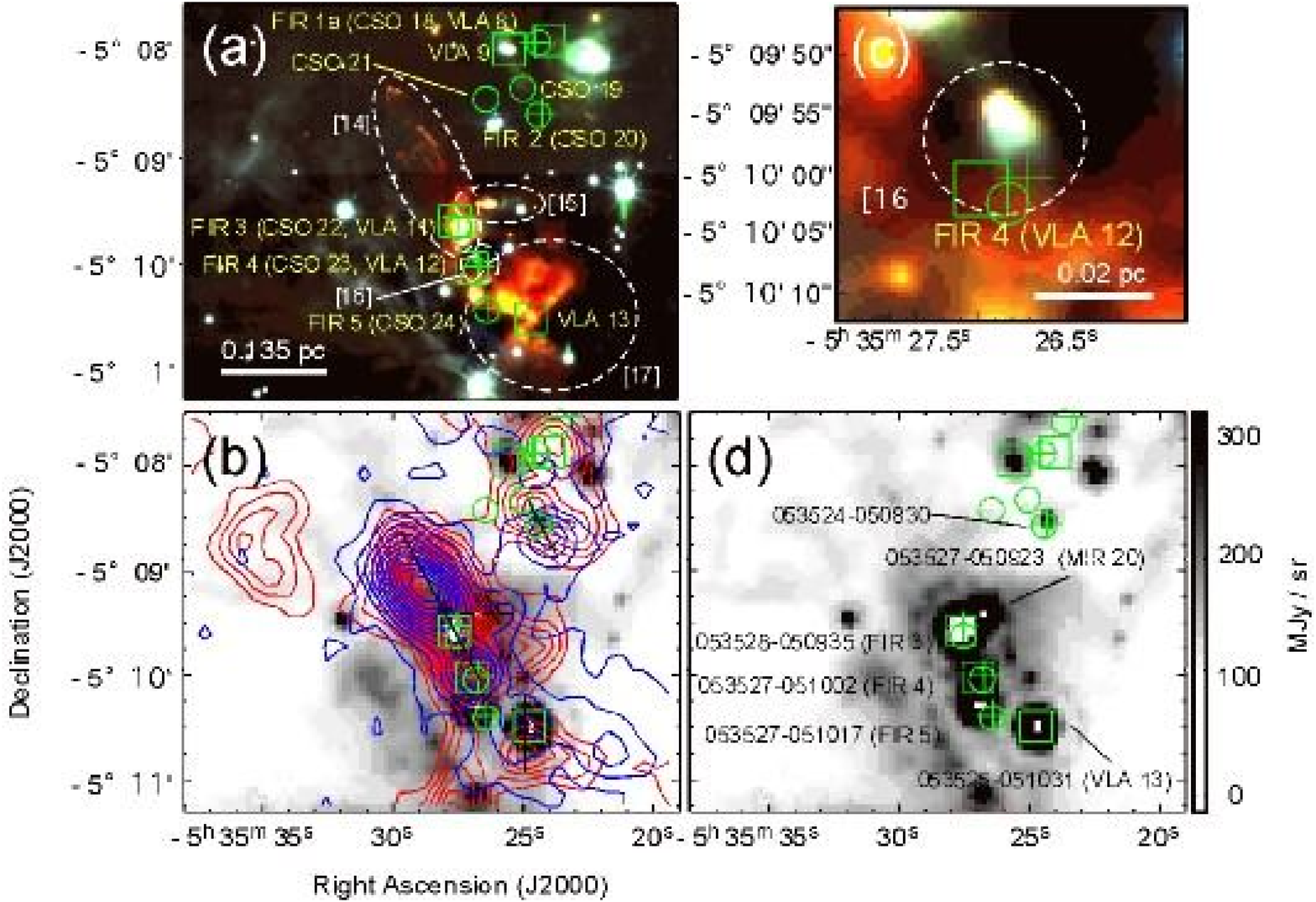}
\caption{
  Images in the FIR 3--5 region. 
  (a) $JHK_s$ image taken by the IRSF/SIRIUS. 
  (b) 24 $\mu$m image taken with Spitzer/MIPS (grey-scale) superposed on 
  the CO (3--2) emission taken with the ASTE telescope (contours).
  The blue and red contours show the blue- and red-shifted components in the velocity range of 
  -4.1 km s$^{-1} \leq V_{\rm{LSR}}~ \leq$  6.7 km$^{-1}$ 
  and 14.3 km s$^{-1}~\leq~V_{\rm{LSR}}~\leq $  21.9 km$^{-1}$, respectively. 
  Blue- and red-contour intervals are 8.40 K km s$^{-1}$ 
  starting at  5.04 K km s$^{-1}$ and 4.62 K km s$^{-1}$ starting at  7.70 K km s$^{-1}$, respectively. 
  (c) Zoom up view of the $JHK_s$ image toward FIR 4.
  (d) The 24 $\mu$m image taken with Spitzer/MIPS (grey-scale) superposed on the source positions.
  Solid- and dashed- lines in panel $b$ show the direction of the outflows categorized as 
  CLEAR (associated with FIR2, FIR 3, and VLA13) and 
  PROBABLE (associated with MIR 20), respectively. 
  Crosses, open circles, and open squares indicate positions of 1.3 mm, 350 $\mu$m and 3.6 cm sources, respectively 
  (from Chini et al. 1977, Lis et al. 1998, Reipurth et al. 1999). 
  Dashed ellipses show the positions of NIR feature presented in Table 2.
  \label{f7}}
\end{figure}

\begin{figure}
\epsscale{1.00}
\plotone{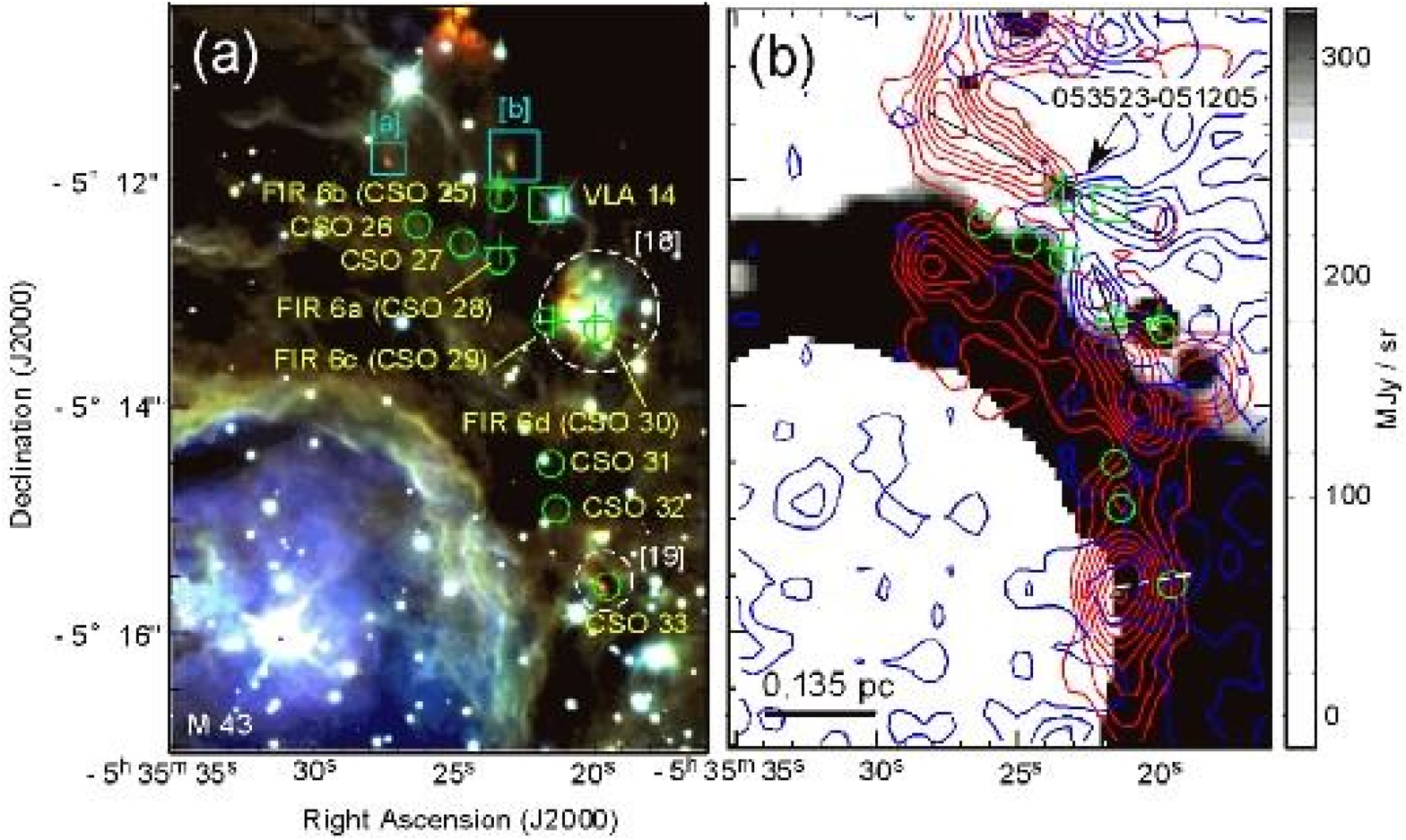}
\caption{
  Images of the southern part of the OMC-2 region. 
  (a) $JHK_s$ image taken by the IRSF/SIRIUS. 
  (b) 24 $\mu$m image taken with SPITZER/MIPS (grey-scale) superposed on 
  the CO (3--2) emission taken with the ASTE telescope (contours).
  The blue and red contours show the blue- and red-shifted components in the velocity range of 
  -3.9 km s$^{-1} \leq V_{\rm{LSR}}~ \leq$  6.7 km$^{-1}$ 
  and 14.3 km s$^{-1}~\leq~V_{\rm{LSR}}~\leq $  21.9 km$^{-1}$, respectively. 
  Contour intervals are 4.65 K km s$^{-1}$ starting at 
  4.65 K km s$^{-1}$ and 4.62 K km s$^{-1}$ starting at  4.62 K km s$^{-1}$, respectively.
  Solid- and dashed- lines in panel $b$ show the outflows categorized as CLEAR (associated with FIR 6$b$) and PROBABLE 
  (associated with FIR 6$c$ and CSO 33), respectively. 
  Crosses, open circles, and open squares indicate positions of 1.3 mm, 350 $\mu$m and 3.6 cm sources, respectively 
  (from Chini et al. 1977, Lis et al. 1998, Reipurth et al. 1999).
  Bright part at bottom left of the figure $a$ is M 43.
  Dashed ellipse shows the position of NIR feature presented in Table 2.
  \label{f8}}
\end{figure}

\begin{figure}
\epsscale{0.80}
\plotone{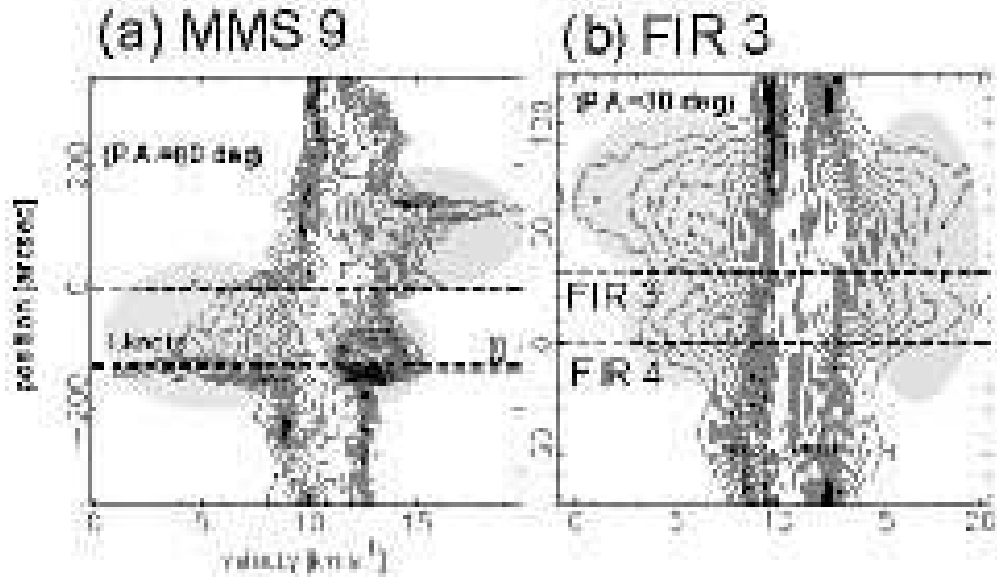}
\caption{Position-velocity diagrams in the CO (3--2) emission along the major axis toward detected outflows in OMC-2/3. 
  The contour starting level and increment are 10\% and 5\% level of the peak intensity, respectively. 
  The reference position of each figure set on the positions of 1.3 mm peak. \label{f9}}
\end{figure}

\begin{figure}
\epsscale{1.00}
\plotone{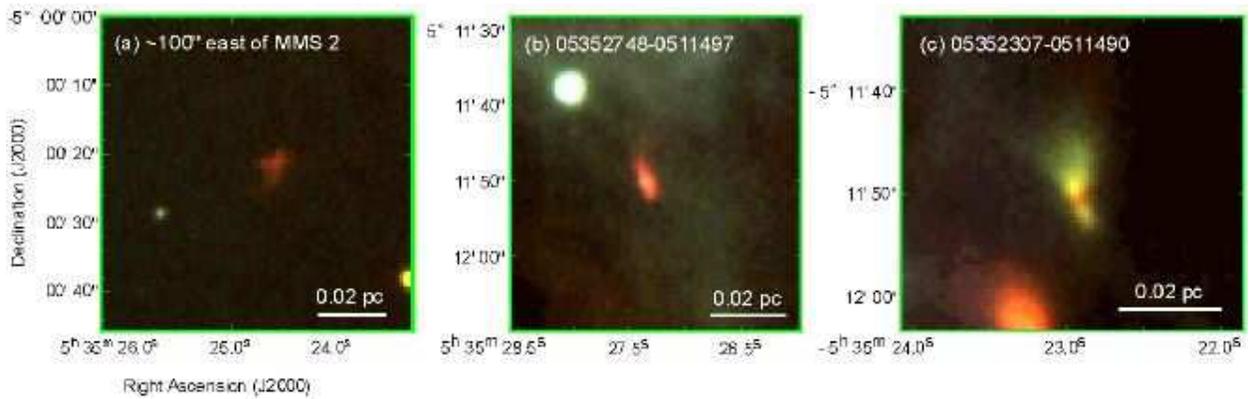}
\caption{$JHK_s$ images of the extended NIR features taken by SIRIUS/IRSF. 
Pseudo colors of blue, green, and red show the data a $J$-, $H$-, and $K$- bands, respectively.\label{f10}}
\end{figure}

\begin{figure}
\epsscale{0.80}
\plotone{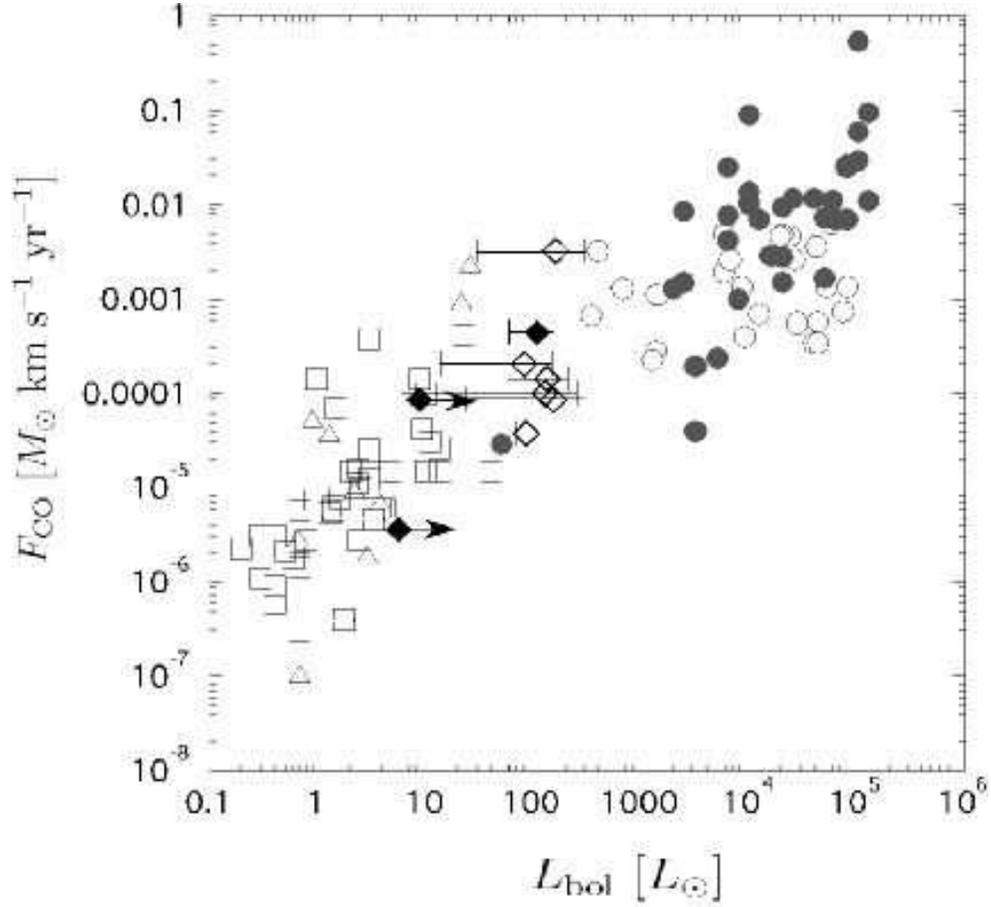}
\caption{CO outflow momentum flux plotted as a function of the bolometric luminosity.
Open and filled diamonds show Class I and Class 0-like protostars in the OMC-2/3 region (this work).
Triangles, open squares, filled circles, and open circles show previous low- to high-mass survey results 
from Hogerheijde (1998), Bontemps et al. (1996), Beuther et al. (2002), and Zhang et al. (2005), respectively.   
\label{f11}}
\end{figure}

\begin{figure}
\epsscale{0.8}
\plotone{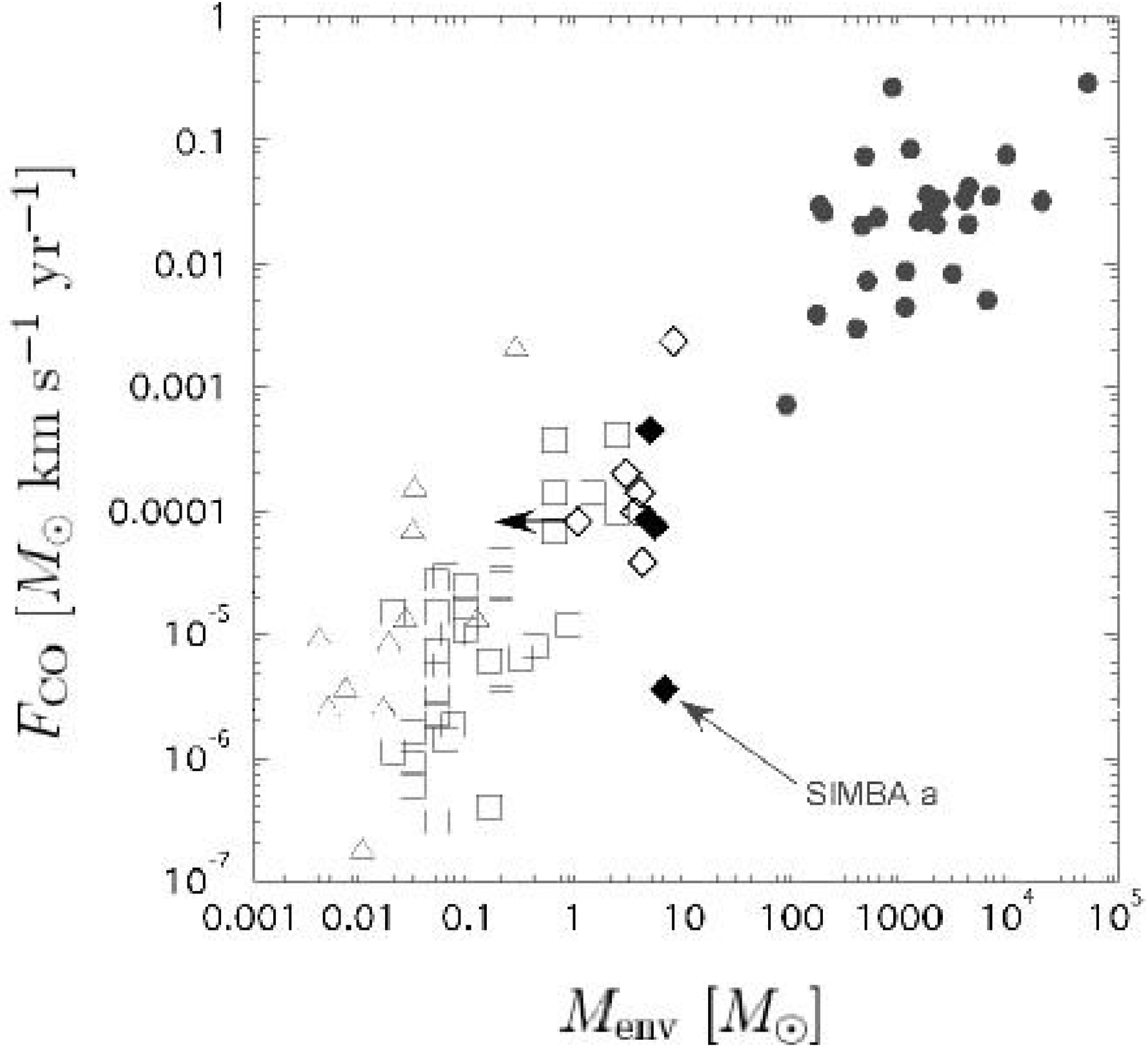}
\caption{CO outflow momentum flux plotted as a function of the envelope mass. 
Open and filled diamonds show Class I and Class 0-like protostars in the OMC-2/3 region (this work). 
Triangles, open squares, and filled circles show previous low- to high-mass survey results 
from Hogerheijde (1998), Bontemps et al. (1996), and Beuther et al. (2002), respectively. 
\label{f12}}
\end{figure}

\clearpage

\begin{deluxetable}{lllccccl}
\tabletypesize{\scriptsize}
\rotate
\tablecaption{Centimeter- and millimeter-/submillimeter- continuum sources in the OMC-2/3 region\label{tbl1}}
\tablewidth{0pt}
\tablehead{
\colhead{ID} & \colhead{R.A. (J2000)} & \colhead{Dec. (J2000)} & \colhead{1.3 mm\tablenotemark{a}} &
\colhead{350 $\mu$m\tablenotemark{b}} & \colhead{3.6 cm\tablenotemark{c}} & \colhead{24 $\mu$m\tablenotemark{d}} &
\colhead{Comments }  \\
}
\startdata
1	& 05 35 30.2	&	-04 58 48.0  	&	SIMBA $a$	&	CSO 1	 &  ---  &	053530-045848 &		 \\
2	&	05 35 29.7 & -04 58 06.0   &	SIMBA $b$ &  	---			   &  ---    & ---	& 	   \\
3	&	05 35 28.2 & -04 58 40.0   &	SIMBA $c$ & 	---			   &  ---     & 053528-045838  &	  \\
4	&	05 35 28.6 & -04 57 48.0   &	SIMBA $d$	& 	---			   & ---	  &  --- 		    &	    \\
5	&	05 35 30.6 & -04 59 49.0   &	SIMBA $e$ & 	---			   &  ---     &  --- 		   &  \\
6	&	05 35 35.1 & -04 59 24.5   &    ---	           &	CSO 2	 & ---		&  ---		  &	 	   \\
7	&	05 35 15.8 & -04 59 58.7   &	---	           &    CSO 3 	 &	---      & 053516-050003 \& 053515-050009	& Binary  		 \\
8	&	05 35 17.1 & -05 00 02.8   &	---	           &    CSO 4	 & ---	 	&  ---		   &	  	\\
9	&	05 35 18.0	& -05 00 19.8	&	MMS 1	 &	   CSO 5	& ---	  &	 ---		  &	        \\
10	&	05 35 18.3	&	-05 00 34.8 &	MMS 2	 &     CSO 6 	&  VLA 1      & 053518-050034		   & Binary (from Nielbock et al. 2003)	 \\
11	&	05 35 19.2 	&	-05 00 51.2	&	MMS 3	 &     CSO 7	&	---           & 053519-050051	 	 	&		\\
12	&	05 35 20.5	&	-05 00 53.0	& 	MMS 4    &     CSO 8	 & ---								    &	 ---		&  \\
13	&	05 35 22.4	&	-05 01 14.1	& 	MMS 5    &	   CSO 9     &	---	          & 053522-050115  	 		&  \\
14	&	05 35 23.5 	&	-05 01 32.2	&	MMS 6    &	   CSO 10	& VLA 3     & 053524-050130  \& 053524-050140				& Multiple system ?  \\
15	&	05 35 27.2  &   -05 03 38.5 &                   &     CSO 11   & ---          						 &  ---             & \\          
16	&	05 35 26.4	&	-05 03 53.4	&	MMS 7    &	   CSO 12	& VLA 4        & 053527-050355				& Bright reflection nebula, Haro- 5a/6a (from Haro 1953). \\
17	&	05 35 26.5	&	-05 05 17.4	&	MMS 8    &	   CSO 13   & --- 								   &  ---			& \\
18	&	05 35 26.0 	&	-05 05 42.4	&	MMS 9    &	   CSO 14	&  VLA 5        &  053526-050546    	 & Extended 24 $\mu$m source. \\
19	&	05 35 32.3 	&	-05 05 41.8	&	MMS 10	&	  ---            & ---								  	&  &  \\
20	&	05 35 25.6	&	-05 07 57.4	&	---			   &	 ---      	    &  VLA 9       & 053526-050758      	&  Located at the edge of the OMC-2 dust filament. \\
21	&	05 35 22.9  &   -05 06 41.2 & 	---			   & CSO 15       &  ---                                  & ---                   & \\ 
22	&	05 35 23.7 	&	-05 07 10.2	&	FIR 1$c$	   &  CSO 16      &  VLA 7           &	---              	                   	&	Extended 24 $\mu$m and cm source.	 \\
23	&	05 35 23.4	&	-05 07 32.2	&   FIR 1$b$	   &  CSO 17	  &	---	                                   &  ---	&		 \\
24	&	05 35 24.6	&	-05 07 53.3	&	FIR 1$a$	   &  CSO 18	  &  VLA 8          & 053524-050753       	& Faint 24 $\mu$m source.	 \\
25	&	05 35 24.7 	&	-05 12 33.3	&	---  	      &  CSO 19	      & ---        & ---	&		 \\
26	&	05 35 24.3 	&	-05 08 33.3	&	FIR 2	    &  CSO 20	    & ---        & 053524-050831        	&		 \\
27	&	05 35 26.5 	&	-05 08 25.4	&	---     	  &  CSO 21	       & ---                                    & ---	&		 \\
28	&	05 35 27.5 	&	-05 09 32.5	&	FIR 3	    &  CSO 22	    & VLA 11     & 053528-050935		    	 & A 24 $\mu$m source (SP 053527-050923) is located \\
		&						&						&					&						&										&	&	at ${\sim}10''$ NW from FIR 3.\\
29	&	05 35 26.7 	&	-05 10 00.4	&	FIR 4	    &  CSO 23	    & VLA 12     & 053527-051002		       & Mm/sub-mm and 24 $\mu$m sources do not \\
		&						&						&					&					&											& 	& coincide with the cm source.	\\
30	&	05 35 24.8 	&	-05 10 29.5	&	---           &  ---				 & VLA 13   & 053525-051031			&  Bright cone-like NIR feature.		 \\
31	&	05 35 26.4 	&	-05 10 23.4	&	FIR 5	   &    CSO 24 	    & ---             & ---	 	   				   &   \\
32	&	05 35 26.3  &   -05 12 23.4  &  ---           &   CSO 26      & ---                                     & ---                 &  \\
33	&	05 35 24.7  &   -05 12 33.3  & ---            &   CSO 27      & ---                                     & ---                 &  \\
34	&	05 35 23.4 	&	-05 12 36.2	&	FIR 6$a$	  &	   CSO 28	  &	--- 									& ---				  &		\\
35	&	05 35 21.7  &   -05 12 13.1 &	---			  &	   ---					  & VLA 14 &	--- 										   &		\\
36	&	05 35 23.4 	&	-05 12 03.2	&	FIR 6$b$	  &	   CSO 25	  &  ---   &	053523-051205         	 			& 	 	\\
37	&	05 35 21.5 	&	-05 13 15.1	&	FIR 6$c$	  &    CSO 29	  &	---                                      &	---					&		\\
38	&	05 35 20.0 	&	-05 13 14.9	&	FIR 6$d$	  &	   CSO 30	  &   ---   &	053520-051314         					&	 \\
39	&	05 35 21.5 	&	-05 14 30.1	&	No obs.	 &	 CSO 31	     & --- 									    &	---					&	 \\
40	&	05 35 21.3 	&	-05 14 54.0	&	No obs.	  &	 CSO 32	     & ---  									&	---					&	 \\
41	&	05 35 19.5 	&	-05 15 34.9	&	No obs.	  &   CSO 33	 &	---	   & 053520-051533	    				& Extended 24 $\mu$m source ?\\
\enddata
\tablenotetext{a}{1.3 mm continuum source names are from Chini et al. (1997) and Nielbock et al. (2003).}
\tablenotetext{b}{350 $\mu$m continuum source names are from Lis et al. (1998).}
\tablenotetext{c}{3.6 cm continuum source names are from Reipurth et al. (1999).}
\tablenotetext{d}{We identified 24 $\mu$m sources obtained from the archive data taken by Spitzer/MIPS using the IRAF apphot package . }
\end{deluxetable}

\clearpage

\begin{deluxetable}{llcccll}
\tabletypesize{\scriptsize}
\rotate
\tablecaption{ IDENTIFICATION OF THE CO (3-2) OUTFLOWS\label{tbl2}}
\tablewidth{0pt}
\tablehead{
\colhead{ID} & \colhead{Object name\tablenotemark{a}} & \colhead{Identification\tablenotemark{b}} & \colhead{Map\tablenotemark{c}} 
& \colhead{NIR features\tablenotemark{d}} & \colhead{Reference\tablenotemark{e}} & \colhead{Comments} \\
}
\startdata
1	   &	SIMBA $a$	&		C	  &	  BR	    &                         & This work                        & \\
3	   &	SIMBA $c$	&		P	  &   BR?	   &		[1] 	       & This work                        &  Faint reflection nebula are associated with blue-shifted CO emission, 
monopolar flow? \\
6	  &	CSO 2	  &		 M	     &   BR	   &		 		                     &  $a$ \& This work               & No Gaussian line profile, \\
	   &				 &				&			 &					                    &                                         & may trace a part of motion of large scale cloud gas \\
10	&	MMS 2	 &  	C	  &   BR	  &	      	  [2], [3], [4]      & $a$, $b$, \& This work  &  Corresponding to ``flow B''\tablenotemark{h}, nearly plane of the sky \\
13	&	MMS 5	 &	    C      &   BR		 &	    	   [5]               & $a$, $b$, \& This work  & Corresponding to ``flow C''\tablenotemark{h}, H$_2$ chain of knots \\
14	&	MMS 6	 & M  &	BR		   &	    [6]                               & This work  & Faint jet-like feature in the NIR image associated with SP 053524-050130 \\
16	&	MMS 7	 &	 	C      &   BR		 &	    	  [7]                &  $a$, $b$, $d$ \& This work & Corresponding to ``flow F''\tablenotemark{h}, nearly plane of the sky\\
18	&	MMS 9	 &		C		&   BR		 &	    	  [9], [10], [11]      & $a$, $b$, \& This work & Corresponding to ``flow H''\tablenotemark{h}\\
20	&	VLA 9	  &		P         &   B		   &	  	[13]                           & This work & Monopolar flow, edge of the dust filament \\
23	&	FIR 1$b$		&	  P         &   BR			 &	                            & $a$, $b$, \& This work  & \\
26	&	FIR 2		 &		C        &   BR		 &	                                       & $b$ \& This work  & \\
28	&	FIR 3		 &		C        &   BR		 &	    [14]                           & $b$, $c$, \& This work & ``Butterfly-like'' velocity structure\\
---\tablenotemark{g}	&	MIR 20     &      M          &  B      &      [15]   & This work   & There is a 24 $\mu$m source, SP 053527-050923 and Bright NIR jet \\
29	&	FIR 4		 &		M       &  BR?		 &	    [16]                           & This work & Faint reflection nebula ?\\
30	&	VLA 13	 &		C          &  BR		  &			[17]                    & This work & Bright NIR feature \\
36	&	FIR 6$b$		&		C        &  BR		&	                               & This work & \\
37	&	FIR 6$c$		&		C         & BR		&	                               & This work  & \\
40	&	CSO 32	  &		 M       &  BR		   &	                                    & This work & Another possibility might be due to gas associated \\ 
		&					&				&				&			                           & This work	&with shocked gas produced by HII region \\
41	&	CSO 33	  &		 P        &  BR		   &	 [19]                            & This work & Another possibility might be due to gas associated \\
		&					&				&				&				&                                         & with shocked gas produced by HII region \\
\enddata
\tablenotetext{a}{Object names from Chini et al. (1997), Nielbock et al. (2003), Lis et al. (1998) and Reipurth et al. (1999)}
\tablenotetext{b}{Outflow identification results based on the CO (3--2) emission and $JHK_s$ image.
C, P, and M indicate outflows identified as CLEAR, PROBABLE, and MARGINAL, respectively.}
\tablenotetext{c}{Bipolarity of the CO (3--2) emission based on channel maps. B and R denote the blue-shifted and red-shifted CO(3--2) emission, respectively.}
\tablenotetext{d}{Notation of near infrared features (see Figure \ref{f2}--\ref{f8})}
\tablenotetext{e}{REFERENCE.---  ($a$)Aso et al. (2000); ($b$)Williams et al. (2003); ($c$)Willams et al. (2005); ($d$)Takahashi et al. (2006).}
\tablenotetext{f}{This outflow corresponds to the [ii] in Figure 6$b$.}
\tablenotetext{g}{CO(3--2) emission and NIR jet-like feature were detected at this source.}
\tablenotetext{h}{Flows were identified in the 2.12 $\mu$m $v$ = 1--0 $s$(1) H$_2$ line emission (Yu et al. 1997)}
\end{deluxetable}

\clearpage

\begin{deluxetable}{llcccccccccc}
\tabletypesize{\scriptsize}
\rotate
\tablecaption{ CO (3-2) OUTFLOW PARAMETERS\label{}}
\tablewidth{0pt}
\tablehead{
\colhead{ID} & \colhead{Object name\tablenotemark{a}} & \colhead{Criteria\tablenotemark{b}} & \colhead{Inclination\tablenotemark{c}} & \colhead{P.A.}  
& \colhead{Velocity Range} &  \colhead{{${\Delta}V_{\rm{max}}$}\tablenotemark{d}} & \colhead{{$R_{\rm{max}}$}\tablenotemark{d}}  &  \colhead{{$M_{\rm{CO}}$}\tablenotemark{e}} 
&  \colhead{{$t_{d}$}\tablenotemark{f}}  &  \colhead{{$\dot{M}_{\rm{CO}}$}\tablenotemark{f}}  &  \colhead{$F_{\rm{CO}}$\tablenotemark{f}} \\ 

\colhead{} & \colhead{} & \colhead{} & \colhead{(deg)} & \colhead{(deg)} & \colhead{(km s$^{-1}$)} & \colhead{(km s$^{-1}$)} & \colhead{(pc)} 
& \colhead{($M_{\odot}$)} & \colhead{(yr)} & \colhead{($M_{\odot}$ yr$^{-1}$)} & 
\colhead{($M_{\odot}$ km s$^{-1}$ yr$^{-1}$)}  \\}

\startdata
\multicolumn{11}{c}{Blue lobe}  \\
\hline
1	&	SIMBA $a$	&	C 	&	45	&	50		& 7.8 -- 8.9 &	4.7	 &	1.2E-01	 &	1.3E-03  & 2.5E+04	& 5.1E-08   & 3.4E-07	 \\
3	&	SIMBA $c$	&	P	&	45	   &	--	   & 8.9 -- 10.0	 &	 3.6  &	1.0E-01 &	1.1E-02  & 2.7E+04	& 4.0E-07	& 2.0E-06  	\\
10	&	MMS 2		  &	  C	  &  45	  &	  90  	  & 3.5 -- 8.9 &	  7.7  &   1.7E-01 &  6.9E-02  & 7.8E+03  & 8.8E-06	  & 2.0E-04	   \\
13	&	MMS 5		  &	  C   &  45	  &	  -90	  & 1.3 -- 8.9 &	  9.7  &   8.4E-02 &  4.0E-02  & 8.5E+03  & 4.7E-06	  & 6.4E-05    \\   
16	&	MMS 7 	 	  &   C	  &	 70	  &	    90     & 4.6 -- 8.9 &  6.0	& 	1.5E-01 &  1.8E-02	& 8.9E+03  & 2.0E-06  & 3.5E-05    \\
18	&	MMS 9         &   C   &	 45	  &		-100  & -4.1 -- 8.9 &	 15.6 &   5.1E-01 &	4.7E-01  & 3.2E+04	 & 1.5E-05	& 3.2E-04   \\
20	&	VLA 9		  &    P   &  45	&	   40	 & 2.4 -- 8.9  &	 9.4	&	3.7E-01	& 1.3E-01 & 3.8E+04   & 3.5E-06	 & 4.6E-05  \\
23	&	FIR 1$b$	&    P   &	45	  &	 	---	   	 &  --- &	 ---	&  ---  & --- & ---   & ---  & --- \\
26	&	FIR 2			&   C	&	45	 &	 	-150  & -4.1 -- 8.9  & 15.4	&  1.0E-01  & 3.2E-02 & 6.3E+03	  & 5.1E-06	 & 1.1E-04 \\	
28	& 	FIR 3 (north)	& C &	70  &	  30	 & -9.5 -- 8.9 &	  20.7	& 1.9E-01  & 1.3E-01 & 3.3E+03	 & 3.9E-05	& 2.3E-03  \\
28	& 	FIR 3 (south)  &  C &   70  &	  30	 & -4.1 -- 8.9 &	  15.3	& 1.0E-01  & 4.8E-02 & 2.3E+03	 & 2.1E-05	& 9.2E-04  \\
30	&	VLA 13	 		 &	C  &   45  &	  0   	 & 2.4 -- 8.9  &	 8.8   & 8.4E-02 &	5.4E-02 & 9.3E+03	& 5.8E-06  & 7.2E-05 \\
36	&	FIR 6$b$		 &	C  &  45   &	   -115 & 1.3 -- 8.9 &	 9.0	& 1.7E-01 &	7.8E-02  & 1.8E+04	 & 4.3E-06	& 5.4E-05 \\
37	&	FIR 6c 			    &  C  &	 45   &	 30		& 0.3 -- 8.9 &	10.0	&  1.2E-01& 2.0E-02	& 1.2E+04	& 1.7E-06  & 2.4E-05  	  \\	 
41	&	CSO 33		   	  &	  P  &  45   &	 	--	  & 3.5 -- 8.9 	 &	 7.2	  &  8.4E-02 & 7.4E-02 & 1.1E+04  & 6.5E-06	& 6.6E-05   \\
\hline
\hline
\multicolumn{11}{c}{Red lobe}  \\
\hline
1	 &   SIMBA $a$		&	C	 & 45		  &	   50	   	  & 13.2 -- 16.5 &	   4.0	    &	1.4E-01	  & 2.0E-02	 & 3.4E+04	& 5.8E-07   & 3.3E-06 	\\
3	 &	 SIMBA $c$		&	P	 & 	45		 	&	 ---	   	 & --- &	   ---	    & 	--- 		   &	---	        & ---	          & ---	           & ---            \\
10	&	MMS 2			  &	  C	   & 45		    &	 90		    & 13.2 -- 16.5  &	 5.3      &	  1.5E-01 & 3.2E-02  & 7.6E+04	 & 4.2E-07	 & 2.4E-06	\\
13	&	MMS 5 	 		  &	  C	   & 45		    &    -90	    & 13.2 --21.9 &   10.9    &	 1.0E-01   & 1.4E-02  & 9.0E+03	 & 1.5E-06	 & 2.3E-05  \\
16	&	MMS 7 	 		  &	  C	   & 70		    &	 90	        & 13.2 -- 18.6 &	8.0		  &    9.8E-01	 & 1.4E-01	& 3.3E+05  & 4.4E-07   & 3.7E-06  \\
18	&	MMS 9 (west) &	 C	  &	45	       &	-100	  & 13.2 -- 26.2 &	  14.7     &   7.1E-01    &	2.8E-01  & 4.7E+04	& 6.0E-06   & 1.2E-04 \\
18	&	MMS 9 (east)  &	  C	   & 45	        &	-100	   & 13.2 -- 17.5 &   6.0 		&	3.5E-01     & 4.9E-02 	& 5.7E+04 & 8.6E-07	  & 7.3E-06 \\
20	&	VLA 9			   &	P    & 45		   &	   40	  & ---  &	 ---	   &	---	  			&	   ---	      & ---			   & ---	       & ---            \\
23	&	FIR 1$b$		  &	   P    & 45		  &	 	--	   	   & 13.2 -- 17.5 &	 6.2	  &  8.4E-02	 &	2.7E-02   & 1.3E+04	& 2.1E-06  & 1.8E-05 \\
26	&	FIR 2				  &	   C    & 45		 &	 -150      & 13.2 -- 20.8 &	 9.5	  &	 6.8E-02      &	1.8E-02   & 7.0E+03 & 2.6E-06  & 3.4E-05 \\
28	&	FIR 3 (north)    &	  C    & 70		  &	 30	            & 13.2 -- 25.1 &	13.9	& 1.9E-01       &  9.7E-02	& 3.7E+04 & 2.7E-06  & 3.9E-05  \\
28	&	FIR 3 (south)    &	  C    & 70		  &	 30	           & 13.2 -- 22.9 &	11.7	& 1.0E-01       & 6.3E-02	& 2.3E+04 & 2.8E-06  & 3.4E-05  \\
30	&	VLA 13	 		   & 	C    & 45		  &	   0           & 13.2 -- 17.5 &	  6.3	   & 6.8E-02	  & 1.4E-02	   & 1.1E+04 & 1.3E-06 & 1.2E-05 \\
36	&	FIR 6$b$		   &	C	 &	45	     &	 -115	    & 13.2 -- 21.9 &	   11.6    & 2.2E-01  &	4.9E-02   & 1.9E+04	 & 2.7E-06 & 4.4E-05    \\
37	&	FIR 6c 			      &	   C	&	45		&	 30	         & 13.2 -- 19.7 &	   9.4	     & 1.8E-01	   &  4.7E-02	& 1.9E+04  & 2.5E-06 & 3.4E-05 \\
41	&	COS 33	   	        &	P	  &	 45		  &	 	---	   	     & --- &	   ---		 & ---	   & ---	& ---  & --- & --- \\
\enddata
\tablenotetext{a}{Object names in the column are from Chini et al. (1997), Lis et al. (1998), Nielbock et al. (2003) and Reipurth et al. (1999).}
\tablenotetext{b}{C and P indicate outflows identified as CLEAR and PROBABLE, respectively.}
\tablenotetext{c}{Inclinations angles of MMS 2, MMS 7, and FIR 3 outflows were assumed nearly plane of the sky 
(i.e., 70$^{\circ}$). Inclination of the other outflows were assumed 45$^{\circ}$. }
\tablenotetext{d}{Not corrected for inclination.}
\tablenotetext{e}{H$_2$ mass of the CO outflow were derived by the LTE assumption. Here, the typical excitation temperature of CO (3--2) 
and the abundance were assumed  $T_{\rm{ex}}=30$ (see Figure \ref{f13}) and $X$[CO]=$10^{-4}$ (Frerking, Langer \& Wilson 1982), 
respectively.}
\tablenotetext{f}{Each parameter assumed inclination angle.}
\end{deluxetable}

\newpage

\begin{table}
  \caption{Physical Properties of the Possible Outflow Driving Sources}\label{t4}
  \begin{center}
    \begin{tabular}{llcccccc}
      \hline
	  ID	& Object name          & $L_{\rm{bol.(min.)}}$ \tablenotemark{a,b}	&	$L_{\rm{bol.(max.)}}$ \tablenotemark{a,c} &	
	  $M_{\rm{env}}$\tablenotemark{d} & Spitzer 24 $\mu$m  & IRAS & Classification\tablenotemark{e}\\
	  		 &									& [$L_{\odot}$]	&	[$L_{\odot}$] &	[$M_{\odot}$]  & & & \\
	  \hline\hline
	  1		&	SIMBA $a$\tablenotemark{f}	          &		5.6	&		---		&	  6.5	   & 053530--045848 &   & 0 \\
	  10		&	MMS 2			  &		14	&		150	        &   	2.8	  & 053518--050034 &  & I \\
	  13		&	MMS 5			  &		9.0	&		---	        &	 4.5	 & 053522--050115 &  & 0 \\
	  16		&	MMS 7			  &		71	&		99		&	   4.1	  & 053527--050355 &  05329--0505 & I \\
	  18		&	MMS 9			  &		60	&		155		&   4.7		 & 053526--050546 & & 0 \\
	  26		&	FIR 2			  &	  59		&		217	        &   3.9	 & 053524--050831 &	 & I  \\
	  28		&	FIR 3			  &	  30		&		300	        &	7.7	  & 053528--050935 & & I \\
	  30		&	VLA 13			  &	  24		&		286	        &	$\leq$ 1.0  & 053525--051031 & & I \\
	  36		&	FIR 6$b$		  &    6.0		&		251	        &	 3.4	& 053523--051205 &  & I \\
	  37		&	FIR 6 $c$		  &    ---\tablenotemark{g}		&		---\tablenotemark{g} 	        &        5.1   & & & 0?     \\
	\hline
	\tablenotetext{a}{The bolometric luminosity of each object is estimated from the total flux integrated  by 8 $\mu$m to 1.3 mm 
	from Chini et al. 1997, Nielbock et al. 2003, and Spitzer archive data.}
	\tablenotetext{b}{Lower limit of the bolometric luminosity. (The data do not include IRAS 100 $\mu$m upper limit fluxes)}
	\tablenotetext{c}{Upper limit of the bolometric luminosity. (The data include IRAS 100 $\mu$m upper limit fluxes)}
	\tablenotetext{d}{Envelope mass were estimated by the flux within the diameter of 5000 AU (11$''$) by Chini et al. (1997).}
	\tablenotetext{e}{Evolutionary status of protostars were classified as Class 0 and Class I based on 2 $\mu$m and 10 $\mu$m results by Nielbock et al. (2003).}
	\tablenotetext{f}{The envelope mass of SIMBA $a$ were estimated by the flux within the diameter of 10800AU (24$''$) by Nielbock et al. (2003)}
	\tablenotetext{g}{The bolometric luminosity was not estimated due to negative detection of the near- to mid-infrared source.}
    \end{tabular}
  \end{center}
\end{table}

\clearpage

\appendix
\begin{figure}
\epsscale{0.80}
\plotone{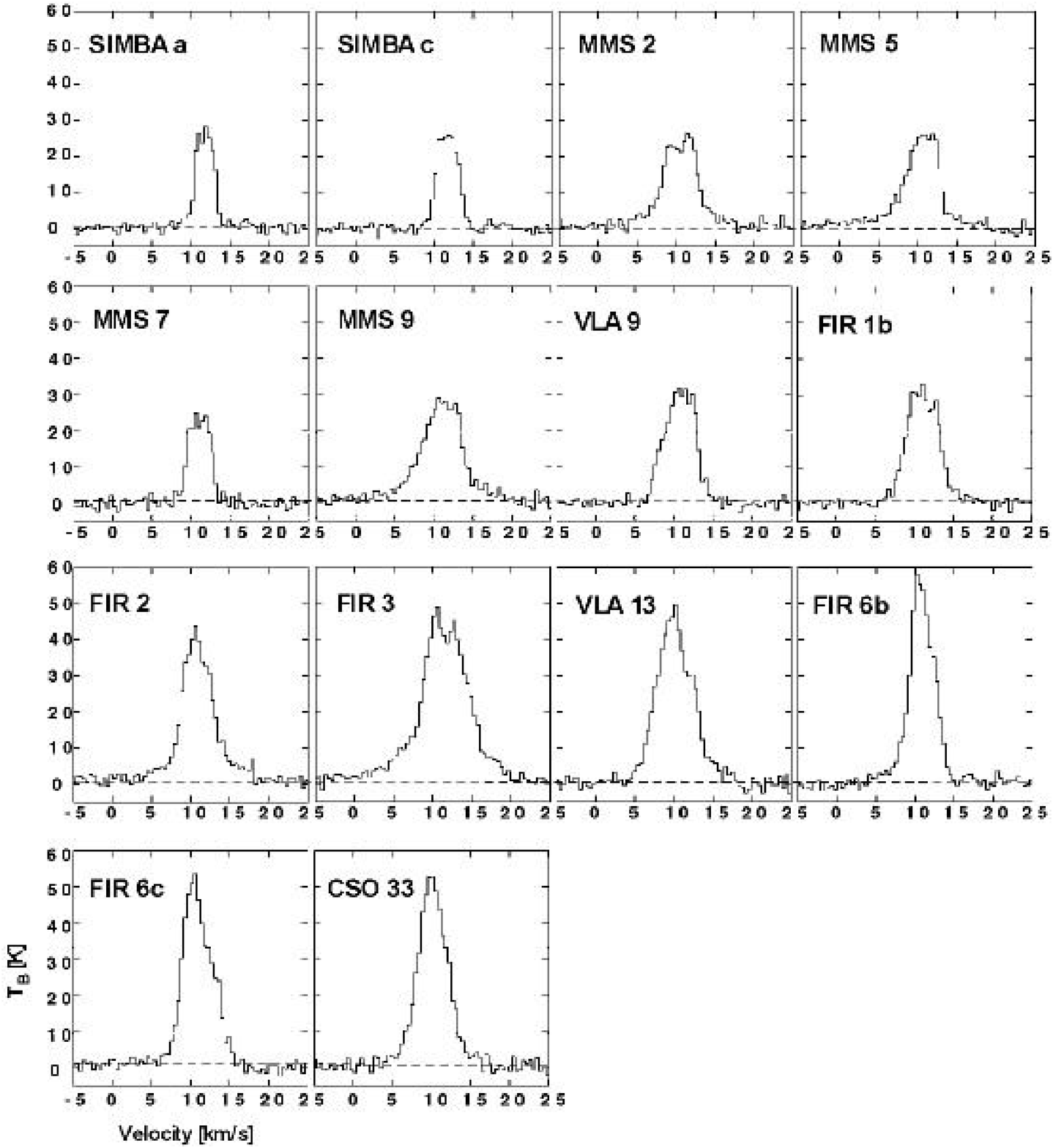}
\caption{CO(3--2) spectra for CLEAR and PROBABLE outflows in our sample at the positions of outflow driving sources 
(i.e., positions of 1.3 mm continuum sources or VLA 3.6 cm sources). \label{f13}}
\end{figure}


\begin{thebibliography}{}
\bibitem[Adams et al (1988)]{ada88} Adams, F. C., Lada, C. J., \& Shu, F. H., 1988, \apj, 326,865 
\bibitem[Andre et al (1993)]{and93} Andre, P., Ward-thompson, D., \& Barsony, M., 1993, \apj, 406, 122 
\bibitem[Allen, L. (2007)]{all07} Allen, L., Megeath, S. T., Gutermuth, R., Myers, P. C., Wolk, S., Adams, F. C., Muzerolle, J., Young, E., \& Pipher, J. L., 
Protostars and Planets V, B. Reipurth, D. Jewitt, and K. Keil (eds.), University of Arizona Press, Tucson, 951pp., 2007., p. 361-376
\bibitem[Arce et al (2006)]{arc06} Arce, H. G., \& Sargent, A. I., 2006, \apj, 646,1070 
\bibitem[Arce et al (2007)]{arc07} Arce, H. G.,  Shepherd, D., Gueth, F., Lee C.-F., Bachiller, R., Rosen, A., \& Beuther, H., 2007, 
Protostars and Planets V (Tucson: Univ. Arizona Press)
\bibitem[Aso. et. al. (2000)]{aso00} Aso, Y., Tatematsu, K., Sekimoto, Y., Nakano, T., Umemoto, T., Koyama, K., \& Yamamoto S., 2000, \apj,	131,	465
\bibitem[Bachiller \& Tafalla (1999)]{bac99} Bachiller, R. \& Tafalla, M., 1999, 
The Origin of Stars and Planetary Systems. Edited by Charles J. Lada and Nikolaos D. Kylafis. Kluwer Academic Publishers, 1999, p.227
\bibitem[Beltran et al. (2006)]{bel06} Beltran, M. T., Girart, J. M., \& Estalella, R., 2006, A\&A, 457, 865
\bibitem[Beltran et al. (2007)]{bel07} Beltran, M. T., Estalella, R., Girart, J. M., Ho, P.T. P., \& Anglada, G., 2007arXiv0712.1757B
\bibitem[Beuther et al. (1992)]{beu92} Beuther, H., Schilke, P., Sridharan, T. K., Menten, K. M., Walmsley, C. M., \& Wyrowski, F.,  2002, A\&A 383,892 
\bibitem[Bontemps et al. (1996)]{bon96} Bontemps, S., Andre, P., Terebey, S., \& Cabrit, S., 1996, A\&A 294, 835
\bibitem[Cabrit et al. (1992)]{cab92} Cabrit, S., \& Bertout, C., 1992, A\&A 261, 274
\bibitem[Cesaroni et al. (1994)]{ces94} Cesaroni, R., \& Wilson, T. L., 1994, A\&A,	281, 209
\bibitem[Chini et al. (1997)]{chi97} Chini, R,.	Ward-Thompson, D,.	Bally, J,.	Nyman,	L.-\AA., Sievers,	A,.	\&	Billawala, Y.,  1997, \apj,	474, L135
\bibitem[Churchwell. (1997)]{chu97} E. Churchwell, 1997, \apj, 479, L59
\bibitem[Emerson \& Graeve (1988)]{eme88} Emerson \& Graeve, 1988, A\&Am 190m 353
\bibitem[Ezawa et al. (2004)]{eza05} Ezawa, H., Kawabe, R., Kohno, K., \& Yamamoto, S., 2004, 
Ground-based Telescopes. Edited by Oschmann, Jacobus M., Jr. Proceedings of the SPIE, Volume 5489, pp. 763
\bibitem[Frerking et al. (1982)]{fre82} Frerking, M., Langer, W. D., \& Wilson, R. W., 1982, \apj, 262, 590 
\bibitem[Fuente et al. (2005)]{fue05} Fuente, A., Neri, R., \& Caselli, P., 2005, A\&A, 448, 481 
\bibitem[Genzel \& Stutzki (1989)]{gen89} Genzel, R. \& Stutzki, J., 1989, ARA\&A, 27, 41
\bibitem[Goldreich \& Kwan (1974)]{gol74} Goldreich, P., \& Kwan, J., 1974, \apj, 189 441
\bibitem[Haro (1953)]{har53} Haro, G. 1953, \apj,	117,	73
\bibitem[Hatchell et al. (2007)]{hat07} Hatchell, J., Fuller, G. A., \& Richer, J. S., 2007, A\&A, 472, 187
\bibitem[Hirano \& Taniguchi (2001)]{hir01} Hirano, N., \& Taniguchi, Y., 2001, \apj, 550, L219 
\bibitem[Hogerheijde et al. (1998)]{hog98} Hogerheijde, M. R., vanDishoeck, E. F., Blake, G. A., \& vanLangevelde, H., J., 1998, \apj, 502, 315
\bibitem[Johnstone et al. (1999)]{joh99} Johnstone, D., \& Bally, J., 1999, \apj, 510, L49
\bibitem[Kamazaki (2005)]{kam05} Kamazaki, T., Ezawa, H., Tatematsu, K., Yamaguchi, N., et al. 2005, 
Astronomical Data Analysis Software and Systems XIV ASP Conference Series, Vol. 347, 
Proceedings of the Conference held 24-27 October, 2004 in pasadena, Calfornia, USA. 
Edited by P. Shopbell, M. Britton, and R. Ebert. San Francisco: Astronomical Society of the Pacific, 2005., p.533
\bibitem[Kohno et al. (2004)]{eza05} Kohno, K., Yamamoto, S., Kawabe, R., Ezawa, H., Sakamoto, S., et al. 2004, 
Springer proceedings in physics, Vol. 91, Berlin, Heidelberg: Springer, 2004, p.349 
Edited by S. Pfalzner, C. Kramer, C. Staubmeier, \& A. Heithausen. 
\bibitem[Lada (1985)]{lad85} Lada, C. J., 1985, ARA\&A, 23, 264
\bibitem[Lada \& Lada (2003)]{lad03} Lada, C. J., \& Lada, E. A., 2003, ARA\&A, 41, 57
\bibitem[Lis et al. (1998)]{lis98} Lis, D.C., Serabyn, E., Keene, J., Dowell, C.D., Benford, D. J., Phillips, T. G., Hunter, T. R., \& Wang, N., 1998, \apj, 509, 299
\bibitem[Mangum et al (2007)]{man07} Mangum, J., Emerson, D., Greisen, E., 2007, A\&A, 474,679 
\bibitem[Matthews et al. (2005)]{mat05} Matthews, B. C., Lai, S. -P., \& Willson, C., 2005, \apj, 626, 959
\bibitem[Motte et al. (1998)]{mot98} Motte, T., Andre, P., \& Neri, R., 1998, A\&A 336, 150 
\bibitem[Nagashima et al. (1999)]{nag99} Nagashima, C., et al. in Star Formation 1999, ed. T. Nakamoto 
(Nobeyama; Nobeyama Radio obs.), 397
\bibitem[Nagayama et al. (2003)]{nag03} Nagayama, T., et al. 2003, Proc. SPIE, 4841, 459
\bibitem[Nakajima et al. (2005)]{nak05} Nakajima, Y., et al. 2005, \aj 129, 776
\bibitem[Nielbock et al. (2003)]{nie03} Nielbock, M., Chini, R., Muller S. A. H., 2003, A\&A, 408, 245
\bibitem[Mutter et al. (2007)]{nu07} Nutter, D., \& Ward-Thompson, 2007, Mon. Not. R. Astron. Soc., 374, 1413
\bibitem[Ohashi et al. (1996)]{oha96}	Ohashi, N,.	Hayashi, M,. Ho, P. T. P., Momose, M., \& Hirano, N., 1996	\apj, 466, 957 
\bibitem[Reipurth et al. (1997)]{rei97} Reipurth, B., Bally, J., \& Devine, D., 1997, \apj,  114, 2708
\bibitem[Reipurth et al. (1999)]{rei99} Reipurth, B., Rodriguea, L. F., \& Chini R., 1999, \apj,  118, 983
\bibitem[Reipurth et al. (2004)]{rei04} Reipurth, B., Rodriguea, L. F., Anglada, G., \& Bally, J., 2004, \apj,  127, 1736
\bibitem[Sawada et al. (2008)]{saw08} Sawada, T et al. (2008) in preparation
\bibitem[Sawada et al. (2008)]{saw08} Sawada, T., Ikeda, N., Sunada, K., \& Kuno, N., et al. 2007arXiv0712.1283s
\bibitem[Shepherd et al. (1996)]{tak96} Shepherd, D. S., \& Churchwell, E., 1996, \apj, 472, 225
\bibitem[Shimajiri et al. (2008)]{shi08} Shimajiri, Y., Takahashi, S., Takakuwa, S., Saito M., \& Kawabe, R., 2008 accepted to \apj.
\bibitem[Shu et al. (1988)]{shu88} Shu, F.H., Lizano, S., Suden S.P., \& Najita J. 1988 \apj 328 L19 
\bibitem[Stanke et al. (2002)]{sta02} Stanke, T., McCaughrean, M. J., Zinnecker H., 2002,  A\&A, 392, 239
\bibitem[Takahashi et al. (2006)]{tak06} Takahashi, S., Saito, M., Takakuwa S., \& Kawabe, R., 2006, \apj, 651, 933
\bibitem[Testi et al. (1998)]{tes98} Testi, L., \& Sargent, A. I., 1998, \apj 508, L91
\bibitem[Tsujimoto et al. (2004)]{tsu04} Tsujimoto, T., Koyama, K., Kobayashi, N., Saito, M., Tsuboi, Y., \& Chandler, C. J. PASJ, 56, 341
\bibitem[Yu et al. (1997)]{yu97} Yu, K. C., Bally, J., \& Devine, D., 1997, \apj, 485, L48
\bibitem[Yu et al. (2000)]{yu00} Yu, K. C., Billawala, Y., Smith, M. D., Bally, J., \& Butner, H. M., 2000, \aj, 120, 1974
\bibitem[Williams et al. (2003)]{wil03} Williams, J. P., Plambeck, R. L., \& Heyer, M. H., 2003, \apj,  591, 1025
\bibitem[Wu et al. (2005)]{wu05} Wu, Y., Zhang, Q., Chen, H., Yang, C., Wei, Y., \& Ho, P.T.P., 2005, \aj, 330, 347
\bibitem[Zhang et al. (2001)]{zha01} Zhang, Q., Hunter, T. R., Brenda, J., Sridharan, T. K., Molinari, S., Kramer, M. A. \& Cesaroni, R., 2001, \apj, 552, L170
\bibitem[Zhang et al. (2005)]{zha05} Zhang, Q., Hunter, T. R., Brenda, J., Sridharan, Cesaroni, R., T. K., Molinari, S., Wang, J., \& Kramer, M. A. \&2005, \apj, 625, 864
\end{thebibliography}
\end{document}